\documentclass[useAMS,usenatbib]{mn2e}

\usepackage{graphicx,natbib}
\usepackage{txfonts}
\usepackage{upgreek}
\usepackage{gensymb}
\usepackage{subfigure}
\newcommand{\msun}{\mbox{M$_{\odot}$}}
\newcommand{\kms}{\mbox{$\rm{km}\,s^{-1}$}}

\def \aj {AJ}
\def \mnras {MNRAS}
\def \apj {ApJ}
\def \apjl {ApJL}
\def \aap {A\&A}
\def \nat {Nature}

\def \pasp {PASP}

\def \apjs {ApJS}

   \title[SN2009ip at late times]{SN 2009ip at late times - an interacting transient at +2 years}

   \author[M. Fraser et al.]
   	{Morgan Fraser,$^{1}$\thanks{E-mail:mf@ast.cam.ac.uk}
    	Rubina Kotak,$^{2}$					
	Andrea Pastorello,$^{3}$				
    	Anders Jerkstrand,$^{2}$\newauthor		
    	Stephen J. Smartt,$^{2}$				
    	Ting-Wan Chen,$^{2}$				
    	Michael Childress,$^{5,6}$			
	Gerard Gilmore,$^{1}$\newauthor 		
  	Cosimo Inserra,$^{2}$				
	Erkki Kankare,$^{2}	$				
	Steve Margheim,$^{4}$				
    	Seppo Mattila,$^{7}$					
    	Stefano Valenti,$^{8,9}$\newauthor		
    	Christopher Ashall,$^{10}$				
	Stefano Benetti,$^{3}$				
	Maria Teresa Botticella,$^{11}$			
    	Franz Erik Bauer,$^{12,13,14}$	\newauthor	
	Heather Campbell,$^{1}$				
    	Nancy Elias-Rosa,$^{3}$				
	Mathilde Fleury,$^{15}$				
    	Avishay~Gal-Yam,$^{16}$	\newauthor		
    	Stephan~Hachinger,$^{17}$ 			
	D. Andrew Howell,$^{9,19}$			
	Laurent Le Guillou,$^{20,15}$\newauthor	
	Pierre-Fran\c{c}ois L\'eget,$^{21}$ 	
    	Antonia~Morales-Garoffolo,$^{22}$		
	Joe Polshaw,$^{2}$					
	Susanna Spiro$^{23},$\newauthor				
    	Mark Sullivan,$^{24}$ 						
    	Stefan Taubenberger,$^{25,26}$				
	Massimo Turatto,$^{3}$				
	Emma S. Walker,$^{27}$\newauthor				
    	David R. Young,$^{2}$					
	Bonnie Zhang$^{5,6}$\\
$^{1}$Institute of Astronomy, University of Cambridge, Madingley Rd., Cambridge, UK\\
$^{2}$Astrophysics Research Centre, School of Mathematics and Physics, Queen's University Belfast, Belfast, BT7 1NN, UK\\
$^{3}$INAF-Osservatorio Astronomico di Padova, Vicolo dell'Osservatorio 5,  35122 Padova, Italy\\
$^{4}$Gemini Observatory, Southern Operations Center, Casilla 603, La Serena, Chile\\
$^{5}$Research School of Astronomy and Astrophysics, The Australian National University, Canberra, ACT 2611, Australia\\
$^{6}$ARC Centre of Excellence for All-sky Astrophysics (CAASTRO), Canberra, ACT 2611, Australia\\
$^{7}$Finnish Centre for Astronomy with ESO (FINCA), University of Turku, V{\"a}is{\"a}l{\"a}ntie 20, FI-21500 Piikki{\"o}, Finland\\
$^{8}$Las Cumbres Observatory Global Telescope Network, 6740 Cortona Dr., Suite 102, Goleta, CA 93117, USA\\
$^{9}$Department of Physics, University of California, Santa Barbara, Broida Hall, Mail Code 9530, Santa Barbara, CA 93106-9530, USA\\
$^{10}$Astrophysics Research Institute, Liverpool John Moores University, IC2, Liverpool Science Park, 146 Brownlow Hill, Liverpool L3 5RF\\
$^{11}$INAF-Osservatorio Astronomico di Capodimonte, Salita Moiariello 16 80131 Napoli Italy\\
$^{12}$Instituto de Astrof\'{\i}sica, Facultad de F\'{i}sica, Pontificia Universidad Cat\'{o}lica de Chile, 306, Santiago 22, Chile\\
$^{13}$Millennium Institute of Astrophysics, Vicu\~{n}a Mackenna 4860, 7820436 Macul, Santiago, Chile\\
$^{14}$Space Science Institute, 4750 Walnut Street, Suite 205, Boulder, Colorado 80301\\
$^{15}$CNRS, UMR 7585, Laboratoire de Physique Nucleaire et des Hautes Energies, 4 place Jussieu, 75005 Paris, France\\
$^{16}$Benoziyo Center for Astrophysics, Weizmann Institute of Science, 76100 Rehovot, Israel\\
$^{17}$Universit\"at W\"urzburg, Lehrstuhl f\"ur Astronomie / Lehrstuhl f\"ur Mathematik IX, Emil-Fischer-Str. 31/30, 97074 W\"urzburg, Germany\\
$^{18}$Institut f\"ur Theoretische Physik und Astrophysik, Universit\"at W\"urzburg, Emil-Fischer-Str. 31, 97074 W\"urzburg, Germany\\.
$^{19}$Department of Astronomy, University of California, Berkeley, CA 94720-3411, USA\\
$^{20}$Sorbonne Universites, UPMC Univ. Paris 06, UMR 7585, LPNHE, F-75005 Paris, France\\
$^{21}$Clermont Universit\'e, Universit\'e Blaise Pascal, CNRS/IN2P3,Laboratoire de Physique Corpusculaire, BP 10448, F-63000 CLERMONT-FERRAND,France\\
$^{22}$Campus UAB, Cam\'i de Can Magrans S/N, 08193 Barcelona, Spain\\
$^{23}$Department of Physics (Astrophysics), University of Oxford, DWB, Keble Road, Oxford OX1 3RH, UK\\
$^{24}$School of Physics and Astronomy, University of Southampton, Southampton, SO17 1BJ, UK\\
$^{25}$European Southern Observatory, Karl-Schwarzschild-Str. 2, 85748 Garching, Germany\\
$^{26}$Max-Planck-Institut f{\"u}r Astrophysik, Karl-Schwarzschild-Str. 1, 85741 Garching bei M{\"u}nchen, Germany\\
$^{27}$Department of Physics, Yale University, PO Box 208120, New Haven, CT 06520-8120, USA\\
}

\begin{document}

\date{Submitted to Monthly Notices of the Royal Astronomical Society}

\pagerange{\pageref{firstpage}--\pageref{lastpage}} \pubyear{}

\maketitle

\label{firstpage}

\clearpage 

\begin{abstract}
We present photometric and spectroscopic observations of the interacting transient SN 2009ip taken during the 2013 and 2014 observing seasons. We characterise the photometric evolution as a steady and smooth decline in all bands, with a decline rate that is slower than expected for a solely $^{56}$Co-powered supernova at late phases. No further outbursts or eruptions were seen over a two year period from 2012 December until 2014 December. SN 2009ip remains brighter than its historic minimum from pre-discovery images. Spectroscopically, SN 2009ip continues to be dominated by strong, narrow ($\lesssim$2000 \kms) emission lines of H, He, Ca, and Fe. While we make tenuous detections of [Fe~{\sc ii}] $\lambda$7155 and [O~{\sc i}] $\lambda\lambda$6300,6364 lines at the end of 2013 June and the start of 2013 October respectively, we see no strong broad nebular emission lines that could point to a core-collapse origin. In general, the lines appear relatively symmetric, with the exception of our final spectrum in 2014 May, when we observe the appearance of a redshifted shoulder of emission at +550 \kms. The lines are not blue-shifted, and we see no significant near- or mid-infrared excess. From the spectroscopic and photometric evolution of SN 2009ip until 820 days after the start of the 2012a event, we still see no conclusive evidence for core-collapse, although whether any such signs could be masked by ongoing interaction is unclear.
\end{abstract}
  



\section{Introduction}
\label{s1}

The physical distinction between Type IIn supernovae (SNe) and SN impostors is clear: Type IIn SNe arise from massive stars\footnote{While some interacting transients have been linked to thermonuclear explosions \protect\citep{Ham03}, we will not discuss these here.} undergoing core-collapse within a dense circumstellar medium (CSM), while SN impostors are eruptions or outbursts from massive stars which do {\it not} undergo core-collapse \citep[][and references therein]{Van00}. Making this distinction from an observational basis is more difficult, as for both Type IIn SNe and SN impostors, the evolution and properties are influenced by the interaction of fast ejecta with the CSM, with similar characteristic narrow line emission in the observed spectra  \citep[e.g.][]{Mau06,Smi11}.

One of the best studied and most controversial SN impostors to date is SN 2009ip. First identified in 2009 by \cite{Maz09}, and initially given a supernova designation\footnote{In this paper, we use the designation ``SN 2009ip'' to refer to the object in question's outburst in 2009, the subsequent period of variability, and the 2012a and 2012b events.}, it was re-classified as an impostor on the basis of its narrow emission lines, low absolute magnitude ({\it V}$\sim$-14) and previous variability \citep{Ber09,Mil09}. Pre-explosion images of NGC 7259 taken with the {\it Hubble Space Telescope} ({\it HST}) revealed the progenitor of SN 2009ip to be a luminous ({\it V}$\sim$-10.3) star with an estimated zero-age main sequence mass of $\sim$60 \msun\ \citep{Smi10,Fol11}. SN 2009ip exhibited sporadic variability over 3 years after discovery, with multiple outbursts reaching absolute magnitudes between {\it R}=-12 and -14 mag \citep{Pas13}. During this period, the spectra were dominated by $\sim$700-1300 \kms\ emission lines of H, along with He, Ca, Na and Fe \citep{Pas13,Smi10,Fol11}. Intriguingly, there was also absorption at velocities of up to 12500 \kms\ as early as September 2011 \citep{Pas13}, implying that at least {\it some} material had been ejected at velocities more usually associated with core-collapse SNe.

In 2012 August, SN 2009ip underwent a new eruption lasting $\sim$40 days, which reached an absolute magnitude of {\it R}=$-$15 mag \citep{Pri13,Pas13,Mau13,Gra14,Mar14}. This eruption was followed by a second outburst $\sim$40 days later, which peaked at an absolute magnitude of {\it R}=$-$18 mag. These two re-brightenings of SN 2009ip were termed the ``2012a'' and ``2012b'' events respectively.  The 2012a event showed a $\sim$20 day rise in all optical bands, followed by a decline on a similar timescale. Broad ($\sim$8000 \kms\ minima) P Cygni lines were seen in H, He and Na, together with strong, narrow emission lines above a blue continuum. Immediately after the 2012a event, SN 2009ip re-brightened dramatically during the 2012b event, increasing by nearly 4 magnitudes in two weeks \citep{Pri13}. The broad absorptions in the spectra of SN 2009ip disappeared at the start of the 2012b event, leaving a blue continuum with narrow, Lorentzian profile lines. Following the maximum of the 2012b event, SN 2009ip underwent a slower decline, which was accompanied by the reappearance of broad absorptions with complex line profiles and multiple absorption components \citep{Fra13,Pas13,Mar14}. \citeauthor{Fra13} and \citeauthor{Mar14} used the magnitude of SN 2009ip at the end of the 2012b event to place a limit of $\lesssim$0.02 \msun\ on the mass of any radioactive $^{56}$Ni synthesised by SN 2009ip.

The CSM surrounding SN 2009ip clearly plays a significant role in the evolution of the transient, and has been studied in detail by multiple groups \citep{Ofe13b,Mar14,Fox15}. \cite{Ofe13b} reviewed all the available mass loss diagnostics for SN 2009ip, including X-rays, radio, and the narrow Balmer lines, and concluded that the average mass-loss rate for the progenitor over the years prior to explosion was between 10$^{-2}$ and $10^{-3}$ \msun\ per year. This value was lower than the mass loss inferred from the H$\alpha$ luminosity of SN 2009ip \citep{Ofe13b,Fra13}, but according to the authors, could be reconciled with it by assuming a non-spherical distribution of CSM. From a detailed study of the Balmer decrement in SN 2009ip, \cite{Lev14} argued that the CSM was likely aspherical, and possibly involved a disk-like structure; while spectropolarimetric observations presented by \cite{Mau14} also indicated a complex geometry for SN 2009ip. \citeauthor{Mau14} suggested that the narrow emission for SN 2009ip arose from an inclined disk, while the broad components came from a distinct, and perpendicular emitting region.

After the 2012a and 2012b events, SN 2009ip appears to be continuing to fade in all bands. The decline rate of SN 2009ip slowed from that initially observed after the peak of the 2012b event, and as of mid 2013 appeared to be approximately half that expected from the decay of radioactive $^{56}$Co \citep{Gra14}. There have been no further reported outbursts or re-brightenings (\citeauthor{Gra14}), however, the magnitude of SN 2009ip still remains brighter than that of the presumed progenitor \citep{Smi10,Fol11}. The late time (post-2012b) spectra of SN 2009ip reveal relatively narrow ($\sim$1-2$\times10^{3}$ \kms) emission from H, He, Na, Ca, and Fe ({\citealp{Gra14,Smi14,Mar14}), albeit without any of the strong nebular lines typically associated with core-collapse SNe. We note, however, that \cite{Fox15} found evidence for very weak [O~{\sc i}] emission in a spectrum obtained at late phases.

Despite the rich observational dataset obtained for SN 2009ip across an exceptionally wide range of wavelengths \citep{Mar14}, the interpretation of the events of late 2012 remains uncertain. During the 2012a and 2012b events, ejecta were seen at velocities of over 10$^4$ \kms\, which together with the unprecedented luminosity, was suggested by \cite{Mau13} to be evidence that SN 2009ip had undergone core-collapse and was now a genuine Type IIn SN. \cite{Smi14} expanded on this scenario, and proposed that the 2012a event was the core-collapse of a blue supergiant, while the 2012b event was powered by the subsequent collision of the ejecta from this event with the surrounding CSM. In contrast, \cite{Pas13}, \cite{Fra13} and \cite{Mar14} remained open as to the final fate of SN 2009ip. \citeauthor{Pas13} argued that as high velocity absorption had already been observed in 2011, this could not be taken as proof positive of core-collapse during either of the 2012a or 2012b events. The absence of emission lines associated with nucleosynthesised elements, and the relatively restrictive limit to the ejected $^{56}$Ni mass could also point to a non-terminal eruption. If not powered by a core-collapse, \citeauthor{Fra13} and \citeauthor{Mar14} suggested that the luminosity of the 2012b event might be explained by the efficient conversion of the kinetic energy of a relatively small eruption as it collided with the CSM. An alternative explanation for SN 2009ip was explored by \cite{Sok13}, \cite{Kas13} and \cite{Tse13}, who proposed that SN 2009ip may be a ``mergerburst'' resulting from two massive stars in a binary system. In such a scenario, the secondary may either merge completely with the primary, or accrete a significant fraction of its mass onto the primary during periastron \citep{Kas13}. The high velocity material seen in SN 2009ip could then be formed by the interaction of a fast jet of material, launched during the merger, with a circumbinary envelope \citep{Tse13}. 

Regardless of whether the 2012a or 2012b events of SN 2009ip were terminal or non-terminal events, the mechanism behind the outbursts seen prior to and including late 2012 remain unknown. There have been several theoretical proposals for increased mass loss shortly before core-collapse caused by vigorous convection during C and later burning stages \citep{Qua12,Smi14b}. Another candidate mechanism is the ``pulsational pair instability'' \citep{Woo07}, which can lead to the release of a large amount of energy, and the ejection of several solar masses of material without destroying the progenitor star. Alternatively, \cite{Sok13} proposed that SN 2009ip could be explained by a mergerburst, where a $\sim$60 \msun\ primary and a lower mass main-sequence secondary merged. In this scenario, the previous outbursts of SN 2009ip during the period 2009 to 2012 August may have been triggered by close periastron passages before the final merger event.

The Public ESO Spectroscopic Survey for Transient Objects \citep[PESSTO;][]{Sma14} observation campaign for SN 2009ip concluded in late 2012 December. In 2013 April, SN 2009ip was visible again from La Silla \citep{Fra13b}; this paper presents observations of SN 2009ip taken during the 2013 observing season until 2013 December, along with observations taken between 2014 April and December. In particular, we address the questions of whether there is new evidence for core-collapse, if there are signatures of dust formation in the ejecta or shocked CSM gas, and crucially, whether SN 2009ip has survived the events of late 2012. The true fate of SN 2009ip is of considerable interest, especially given the recent identification of pre-explosion outbursts in several SNe \cite[e.g][]{Pas07,Ofe13a,Fra13c,Ofe14}. SN 2009ip is also one of the best monitored SNe of any type, with 5 years of data since its discovery, and as such represents a unique window into the lives of massive stars. 

In all of the following we adopt a distance modulus of 31.55 mag, and a foreground extinction of $A_R$=0.051 mag towards SN 2009ip, consistent with that used by \cite{Smi10,Pas13,Fra13}. We also adopt 2012 September 23 (MJD 56193) as the start of the 2012b event, all phases quoted in the paper are with respect to this date.

\section{Observations and data reduction}
\label{s2}

\subsection{Imaging}

Optical imaging of SN 2009ip was obtained with the 3.6m New Technology Telescope (NTT) + ESO Faint Object Spectrograph and Camera 2 (EFOSC2), the 1.3 m Small and Moderate Aperture Research Telescope System (SMARTS) telescope + ANDICAM (operated by the SMARTS Consortium), the 2 m Faulkes South telescope + Spectral imager (FS03) and the 2 m Liverpool Telescope + Infrared-Optical:Optical (IO:O) camera. The EFOSC2 images were reduced using the PESSTO pipeline, as described in \cite{Sma14}, while the ANDICAM, FS03 and IO:O images were automatically reduced with their associated instrument pipelines. In addition to these, we used the filtered acquisition images taken before our Very Large Telescope (VLT) + XShooter and Gemini South + Gemini Multi-Object Spectrograph (GMOS) spectra. The XShooter acquisition image is taken with a separate acquisition camera to the main instrument, for which biases and flat-fields are not available. However, the image did not show a strong gradient in the background, and so we assume the flat field correction is small. The Gemini images were reduced using the tasks in the {\sc gemini} package within {\sc iraf}. The ANDICAM, IO:O, FS03 and EFOSC2 images were processed with {\sc lacosmic} to remove cosmic rays, before consecutive exposures taken with the same instrument and filter were coadded. As only single, short exposures were taken with GMOS and the XShooter acquisition camera, these data were not coadded or processed with {\sc lacosmic}.

For all images, point-spread function (PSF) fitting photometry was carried out using {\sc snoopy}, a custom {\sc iraf} package based on {\sc daophot}. The PSF was modelled using point sources on each frame, and was fit simultaneously to both SN 2009ip and the nearby foreground star at ($\sim$5\arcsec to the NE at 22\textsuperscript{h}23\textsuperscript{m}5\fs51 $-28\degree56\arcmin47\farcs9$) to account for any additional flux from the latter. The fit was iterated to minimise the residual, while artificial star tests at the SN position were used to estimate the uncertainty in magnitude. The zeropoint for each frame was determined using aperture photometry of the tertiary sequence stars listed by \cite{Pas13}, with the uncertainty in the zeropoint taken to be the standard deviation of the values calculated from each of the individual sequence stars. Aperture corrections were also measured from the sequence stars, and where necessary applied to the photometry of SN 2009ip, and colour terms were applied to the measured magnitudes as in \cite{Fra13}.

At these late epochs, the spectrum of SN 2009ip is dominated by line rather than continuum emission. As a consequence of this, the photometry of SN 2009ip is sensitive to the precise transmission of each filter used. As shown in Fig. \ref{fig:filters}, the Bessell and Sloan filters from the various instruments used for observations of SN 2009ip have somewhat different transmissions. These differences result in flux from various lines being included or excluded from a particular passband -- for example the EFOSC2 {\it i} filter includes flux from the $\lambda$8662 line of the Ca~{\sc II} NIR triplet, while the IO:O {\it i} filter does not. As a check on the significance of the flux present in line emission at this epoch, we took the GMOS spectrum from August 12 and set the flux in the region of H$\alpha$ to the continuum level. Synthetic Bessell {\it R}-band photometry performed with {\sc iraf synphot} on both the original and H$\alpha$-free spectra gives a magnitude for the latter which is 1.3 mag fainter, demonstrating line flux can have a significant effect on our measured photometry.

It is possible to ameliorate this to some extent using synthetic photometry of spectra \citep[``S-correction'', e.g.][]{Str02}, however this is dependent on the availability of carefully measured total transmission functions (i.e. telescope optics, filter and CCD), which were unavailable for several of the instruments we have used. Hence we have not performed a full S-correction for all filters, but rather have applied a simple correction using the available filter functions, for the instruments and filters where there was the greatest disagreement. We have also taken the EFOSC2 filters as our reference filter system, rather than the standard \cite{Bes90} filter curves, with the motivation of both reducing the number of corrections which we make to our data (i.e. the EFOSC2 data does not have to be corrected), and also to ensure consistency with the data presented in \cite{Fra13}. Average colour terms to convert EFOSC2 photometry to other systems are available in \cite{Sma14}.

For each photometric point to be corrected, we have taken the closest available spectrum of SN 2009ip, and with {\sc synphot}, measured the difference in synthetic magnitude when using the specific filter function for that instrument, and when using the corresponding EFOSC2 filter ($\Delta M^{SN}_{INS-EFOSC2}$). This difference cannot be applied directly to the SN 2009ip magnitudes, as some fraction of it will already be accounted for by the applied colour term. To estimate the portion of this difference which is accounted for by the colour term, we first estimated the effective temperature of the SN, using a blackbody fit to the spectrum (excluding the regions around H$\alpha$ and H$\beta$). We then took a Kurucz model at a similar temperature, and appropriate to a solar metallicity dwarf, on which we also measured the difference in synthetic magnitude ($\Delta M^{*}_{INS-EFOSC2}$) using the filter functions for the instrument in question, and the filter function for EFOSC2. This difference is what one would expect for a continuum-dominated source, such as those stars from which the applied colour terms are derived. Finally, we can calculate $\left(\Delta M^{SN}_{INS-EFOSC2}\right)-\left(\Delta M^{*}_{INS-EFOSC2}\right)$, which is the additional correction to the measured photometry of the SN at that epoch (in addition to the colour term). The final calibrated magnitudes for SN 2009ip in the EFOSC2 ($\sim$Landolt) system, together with the total uncertainty (comprised of the error in SN magnitude and zeropoint, added in quadrature) are reported in Table \ref{tab:phot}. For the LT+IO:O images which were taken in Sloan-like {\it r} and {\it i}-filters, we were unable to homogenise the measured photometry with the data from other telescopes, even after applying the correction from synthetic photometry. We have hence excluded these images from our analysis. With the exception of the LT {\it ri} images, the measured magnitudes for SN 2009ip using different telescopes on the same night are consistent to within the uncertainties.

\begin{figure}
\centering
\includegraphics[width=0.9\linewidth,angle=0]{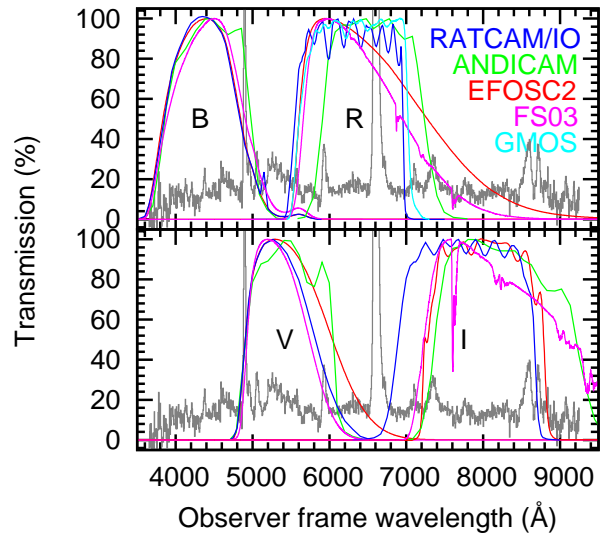}
\caption{Filter functions for the filters used for the followup of SN 2009ip; for FS03 these are transmission functions which also include the contribution of the Earths atmosphere. All transmission filters have been normalised to 100\% transmission at their peak. For comparison, an observer frame spectrum of SN 2009ip from 2013 April is shown in grey.}
\label{fig:filters}
\end{figure}

\begin{table*}
\caption{Calibrated optical magnitudes of SN 2009ip in the Landolt system, with associated uncertainties in parentheses. Phase is with respect to the start of the 2012a event (MJD 56193).}
\begin{center}
\begin{tabular}{lllccccr}
\hline
Phase (d)	& Date 		& MJD	         & {\it B} (mag)			& {\it V} (mag)			& {\it R} (mag)			& {\it I} (mag)			& Instrument 	\\
\hline
192	&   2013-04-03.40	&   56385.40	 &   -  	    & 19.23   (0.06) &  18.24  (0.06) & -		&    EFOSC2	  \\		      
193	&   2013-04-04.38	&   56386.38	 &   -  	    & 19.25   (0.05) &  -	      & -		&    EFOSC2	  \\	 	      
195	&   2013-04-06.41	&   56388.41	 &   19.95   (0.10) & - 	     &  -	      & -		&    EFOSC2	  \\
197	&   2013-04-08.40	&   56390.40	 &   -  	    & 19.26   (0.10) &  18.32  (0.07) & 18.86	(0.08)  &    ANDICAM	  \\	 	       
201	&   2013-04-12.39	&   56394.39	 &   -  	    & 19.43   (0.07) &  18.42  (0.08) & 18.76	(0.10)  &    ANDICAM	  \\	 	       
202	&   2013-04-13.37	&   56395.37	 &   -  	    & 19.25   (0.06) &  -	      & -		&    EFOSC2	  \\	 	       
203	&   2013-04-14.37	&   56396.37	 &   -  	    & 19.34   (0.07) &  -	      & -		&    EFOSC2	  \\	 	       
205	&   2013-04-15.90	&   56397.90	 &   -  	    & 19.34   (0.11) &  18.29  (0.06) & 18.80	(0.09)  &    ANDICAM	  \\	 	       
209	&   2013-04-20.37	&   56402.37	 &   -  	    & 19.38   (0.05) &  -	      & -		&    EFOSC2	  \\	 	       
219	&   2013-04-29.85	&   56411.85	 &   -  	    & - 	     &  18.47  (0.05) & -		&    ANDICAM	  \\	 	       
222	&   2013-05-03.37	&   56415.37	 &   -  	    & - 	     &  18.46  (0.12) & -		&    GMOS	  \\	 	      
225	&   2013-05-06.33	&   56418.33	 &   -  	    & - 	     &  18.42  (0.06) & -		&    XShooter	  \\	 	      
239	&   2013-05-20.34	&   56432.34	 &   -  	    & - 	     &  18.55  (0.06) & -		&    GMOS	  \\	 	      
242	&   2013-05-23.40	&   56435.40	 &   -  	    & - 	     &  18.59  (0.05) & -		&    GMOS	  \\	      
245	&   2013-05-25.78	&   56437.78	 &   20.48   (0.09) & 19.89   (0.06) &  18.53  (0.05) & 18.96	(0.08)  &    FS03	  \\	      
248	&   2013-05-28.79	&   56440.79	 &   -  	    & 19.93   (0.11) &  18.61  (0.07) & -		&    FS03	  \\	      
254	&   2013-06-03.69	&   56446.69	 &   20.55   (0.05) & 20.02   (0.07) &  18.62  (0.04) & 19.12	(0.08)  &    FS03	  \\	      
254	&   2013-06-04.34	&   56447.34	 &   -  	    & - 	     &  18.67  (0.03) & -		&    GMOS	  \\	      
264	&   2013-06-14.35	&   56457.35	 &   -  	    & - 	     &  18.77  (0.05) & -		&    GMOS	  \\	 	      
269	&   2013-06-18.82	&   56461.82	 &   20.34   (0.05) & 20.08   (0.05) &  18.70  (0.04) & 19.22	(0.04)  &    FS03	  \\	      
280	&   2013-06-30.40	&   56473.40	 &   -  	    & - 	     &  18.82  (0.03) & -		&    GMOS	  \\	 	      
293	&   2013-07-12.81	&   56485.81	 &   20.73   (0.04) & 20.17   (0.07) &  18.76  (0.04) & 19.34	(0.08)  &    FS03	  \\	      
303	&   2013-07-23.33	&   56496.33	 &   -  	    & - 	     &  18.85  (0.05) & -		&    GMOS	  \\	 	      
306	&   2013-07-26.11	&   56499.11	 &   20.83   (0.23) & 20.02   (0.10) &  -	      & -		&    IO:O	  \\	 	       
308	&   2013-07-27.70	&   56500.70	 &   20.67   (0.07) & 20.12   (0.06) &  18.82  (0.04) & -		&    FS03	  \\    
308	&   2013-07-28.34	&   56501.34	 &   20.68   (0.08) & 20.06   (0.05) &  18.97  (0.06) & 19.27	(0.05)  &    EFOSC2	  \\	 	       
311	&   2013-07-30.70	&   56503.70	 &   20.91   (0.11) & 20.36   (0.10) &  18.89  (0.06) & 19.58	(0.13)  &    FS03	  \\    
316	&   2013-08-05.20	&   56509.20	 &   20.71   (0.06) & 20.08   (0.05) &  19.01  (0.06) & 19.45	(0.06)  &    EFOSC2	  \\	 	       
317	&   2013-08-05.79	&   56509.79	 &   20.70   (0.05) & 20.14   (0.08) &  18.82  (0.04) & 19.44	(0.10)  &    FS03	  \\    
319	&   2013-08-08.02	&   56512.02	 &   20.67   (0.09) & 20.02   (0.09) &  -	      & -		&    IO:O	  \\	 	      
323	&   2013-08-11.62	&   56515.62	 &   20.78   (0.08) & 20.27   (0.07) &  18.90  (0.04) & -		&    FS03	  \\	      
323	&   2013-08-12.15	&   56516.15	 &   -  	    & - 	     &  18.93  (0.12) & -		&    GMOS	  \\	 	      
326	&   2013-08-15.13	&   56519.13	 &   20.82   (0.04) & 20.20   (0.05) &  19.03  (0.07) & 19.52	(0.05)  &    EFOSC2	  \\	 	       
327	&   2013-08-16.07	&   56520.07	 &   20.88   (0.12) & 20.06   (0.05) &  -	      & -		&    IO:O	  \\	 	       
329	&   2013-08-17.59	&   56521.59	 &   20.88   (0.09) & 20.39   (0.05) &  18.96  (0.03) & 19.59	(0.10)  &    FS03	  \\    
338	&   2013-08-26.63	&   56530.63	 &   -  	    & 20.20   (0.19) &  18.99  (0.12) & -		&    FS03	  \\    
338	&   2013-08-27.10	&   56531.10	 &   20.88   (0.06) & 20.27   (0.06) &  19.14  (0.06) & 19.61	(0.09)  &    EFOSC2	  \\	 	       
341	&   2013-08-29.62	&   56533.62	 &   -  	    & - 	     &  19.01  (0.08) & -		&    FS03	  \\    
341	&   2013-08-30.01	&   56534.01	 &   20.96   (0.12) & 20.09   (0.07) &  -	      & -		&    IO:O	  \\	 	       
351	&   2013-09-09.04	&   56544.04	 &   -  	    & 20.17   (0.07) &  -	      & -		&    IO:O	  \\	 	       
354	&   2013-09-12.20	&   56547.20	 &   20.75   (0.05) & 20.14   (0.04) &  19.12  (0.05) & 19.66	(0.08)  &    EFOSC2	  \\	 	       
361	&   2013-09-18.95	&   56553.95	 &   -  	    & 20.24   (0.23) &  -	      & -		&    IO:O  \\	 	       
369	&   2013-09-26.95	&   56561.95	 &   20.77   (0.08) & 20.26   (0.07) &  -	      & -		&    IO:O	  \\	 	       
376	&   2013-10-04.05	&   56569.05	 &   20.99   (0.06) & 20.39   (0.10) &  19.23  (0.04) & 19.77	(0.07)  &    EFOSC2	  \\	 	       
388	&   2013-10-16.18	&   56581.18	 &   20.86   (0.13) & 20.46   (0.15) &  19.30  (0.06) & 19.82	(0.08)  &    EFOSC2	  \\	 	       
397	&   2013-10-25.14	&   56590.14	 &   21.07   (0.10) & 20.52   (0.11) &  19.34  (0.08) & 19.96	(0.24)  &    EFOSC2	  \\	 	       
409	&   2013-11-05.83	&   56601.83	 &   -  	    & 20.41   (0.32) &  -	      & -		&    IO:O	  \\	 	       
427	&   2013-11-23.87	&   56619.87	 &   21.24   (0.09) & 20.51   (0.07) &  -	      & -		&    IO:O	  \\	 	       
430	&   2013-11-26.82	&   56622.82	 &   21.25   (0.13) & 20.50   (0.07) &  -	      & -		&    IO:O	  \\	 	       
434	&   2013-12-01.05	&   56627.05	 &   21.19   (0.06) & 20.67   (0.05) &  19.43  (0.04) & 19.98	(0.09)  &    EFOSC2	  \\	 	       
457	&   2013-12-24.05	&   56650.05	 &   21.25   (0.07) & 20.71   (0.04) &  19.45  (0.05) & 20.13	(0.05)  &    EFOSC2	  \\	 	       
575	&   2014-04-20.32	&   56768.32	 &   -			   & 21.05	 (0.05)  &	19.80  (0.07)	& 20.40  (0.27)   &    EFOSC2	  \\
577	&   2014-04-22.36	&   56770.36	 &   21.62	 (0.10) &	-		    &	 -	 		& -			 &     EFOSC2	  \\
578	&   2014-04-23.40	&   56771.40	 &   -			   & 21.17 	 (0.06)  &	19.84  (0.05)	& 20.65  (0.10)  &    EFOSC2	  \\
583	&   2014-04-28.38	&   56776.38	 &   - 		   & 21.03	 (0.18)  &	19.80  (0.11)	& 20.52  (0.21)	 &    EFOSC2	  \\
589	&   2014-05-04.37	&   56782.37	 &   21.61	 (0.61) & 21.10	 (0.09)  &	19.90  (0.08)	& 20.83  (0.29)	 &  EFOSC2	  \\
692	&   2014-08-16.12	&   56885.12	 & 21.92	(0.08)    & 21.45 (0.05)	& 20.21	(0.04) & 20.89	(0.08) & EFOSC2 \\	
723	&   2014-09-16.20	&   56916.20	 &  -		     &  -		        &19.90	(0.08) & -			   & EFOSC2 \\ 
731	&   2014-09-24.02	&   56924.02	 &  21.79 (0.13)    & 21.45 (0.05)	& 20.27	(0.06) & -			   &  EFOSC2 \\	
736	&   2014-09-29.21	&   56929.21	 &  21.94 (0.10)    & 21.54 (0.07)	& 20.28	(0.05) & -			   &  EFOSC2 \\
782 	&   2014-11-14.06	&   56975.06	 &  21.99	(0.09)    & 21.61 (0.09)      & 20.44 (0.06)    & 21.21 (0.08)   & EFOSC2 \\
821 	&   2014-12-23.06	&   57014.06	 & 22.00 (0.07)   & 21.73 (0.07)			& 20.55 (0.07)	& 21.45 (0.31) 	       & EFOSC2 \\
\hline
\end{tabular}
\end{center}
\label{tab:phot}
\end{table*}%

We also took near-infrared (NIR) imaging of SN 2009ip in {\it JHKs} filters using both the NTT + Son of ISAAC (SOFI) and the 2.6m Nordic Optical Telescope + NOTCam. The SOFI data were reduced using the PESSTO pipeline \citep{Sma14}, while the NOTCam images were reduced using the external {\sc notcam} package (version 2.5) within {\sc iraf}. In both cases, the images were flat-fielded, and sky-subtracted using a sky frame constructed from median combined on-source dithers. PSF-fitting photometry was performed on the SN using the same procedure as for the optical imaging, although the zeropoint was determined directly from catalogued Two Micron All-Sky Survey (2MASS) sources in the field, and no colour terms were applied. The calibrated magnitudes are reported in Table \ref{tab:irphot}. For epochs where SN 2009ip could not be seen in these images, a limiting magnitude was estimated from artificial star tests at the expected position of the transient.

Mid-infrared (MIR) imaging was taken of SN 2009ip at multiple epochs in 2013 with the {\it Spitzer Space Telescope} + Infrared Array Camera (IRAC), as listed in Table \ref{tab:irphot}. As {\it Spitzer} is now in its warm mission phase, SN 2009ip was observed only in Channel 1 (3.6$\upmu$m) and Channel 2 (4.5$\upmu$m). Images from the {\it Spitzer} archive taken in 2009 and 2010 were also examined, but the position of the transient lies off the edge of the detector in these data, and hence these were not considered any further. Photometry was performed on the post-Basic Calibrated Data (pBCD) images using the {\sc mopex} package \citep{Mak05}. A mosaic of the field was created (an example of which is shown in Fig. \ref{fig:mir}), and sources were identified on this image, before Point Response Function (PRF) fitting photometry was performed on the individual frames. Measured fluxes were then corrected for normalisation as per the IRAC Instrument Handbook. No array-position dependent colour correction was made to the data, although as the SN is not particularly blue at these late epochs, the effect of this should be negligible. The measured fluxes for SN 2009ip are listed in Table \ref{tab:irphot}.

\begin{table*}
\caption{Calibrated NIR magnitudes and MIR fluxes for SN 2009ip.}
\begin{center}
\begin{tabular}{llccccccr}
\hline
Phase (d) 	& Date 		& MJD		& {\it J} (mag)		& {\it H} (mag)		& {\it K} (mag)		& 3.6$\upmu$m (mJy)	& 4.5$\upmu$m (mJy) 	& Instrument 	\\
\hline
130		& 2013-01-31.01	& 56323.01	& -			& -			& -			& 0.1331   (0.0013)	& 0.1628   (0.0012)   	& IRAC		\\
201		& 2013-04-11.37	& 56394.37	& 17.94 (0.06)	& 17.92 (0.16)	& 17.83 (0.11)	& -					& -					& SOFI		\\
303		& 2013-07-23.16	& 56496.16	& 18.84 (0.18)	& 18.99 (0.20)	& 18.12 (0.17)	& -					& -					& NOTCam	\\
306		& 2013-07-25.95	& 56498.95	& -			& -			& -			& 0.0311   (0.0012)	& 0.0460   (0.0009)   	& IRAC		\\
326		& 2013-08-14.55	& 56518.55	& -			& -			& -			& 0.0286   (0.0011)	& 0.0464   (0.0008) 	& IRAC		\\
339		& 2013-08-27.74	& 56531.74	& -			& -			& -			& 0.0255   (0.0011)	& 0.0474   (0.0008)  	& IRAC	 	\\
341		& 2013-08-30.02	& 56534.02	& 19.05 (0.16)	& 19.41 (0.26)	& 18.53 (0.24)	& -					& -					& NOTCam	\\
364		& 2013-09-21.98	& 56556.98	& 19.44 (0.19)	& 19.36 (0.26)	& $<$18.41	& -					& -					& NOTCam	\\	
376		& 2013-10-03.22	& 56569.22	& 19.18 (0.12) 	& -			& -			& -					& -					& SOFI		\\
428		& 2013-11-24.77	& 56620.77	& 19.68 (0.23)	& $<$18.84	& -			& -					& -					& NOTCam	\\
711		& 2014-09-03.07	& 56904.07	& 20.67 (0.29)	& $<$20.5		& $<$18.8		& -					& -					& SOFI		\\
730		& 2014-09-22.09	& 56923.09	& 20.78 (0.21)	& -			& -			& -					& -					& SOFI		\\
\hline
\end{tabular}
\end{center}
\label{tab:irphot}
\end{table*}%

\begin{figure}
\subfigure[3.6$\upmu$m]{
\includegraphics[height=41mm,width=41mm]{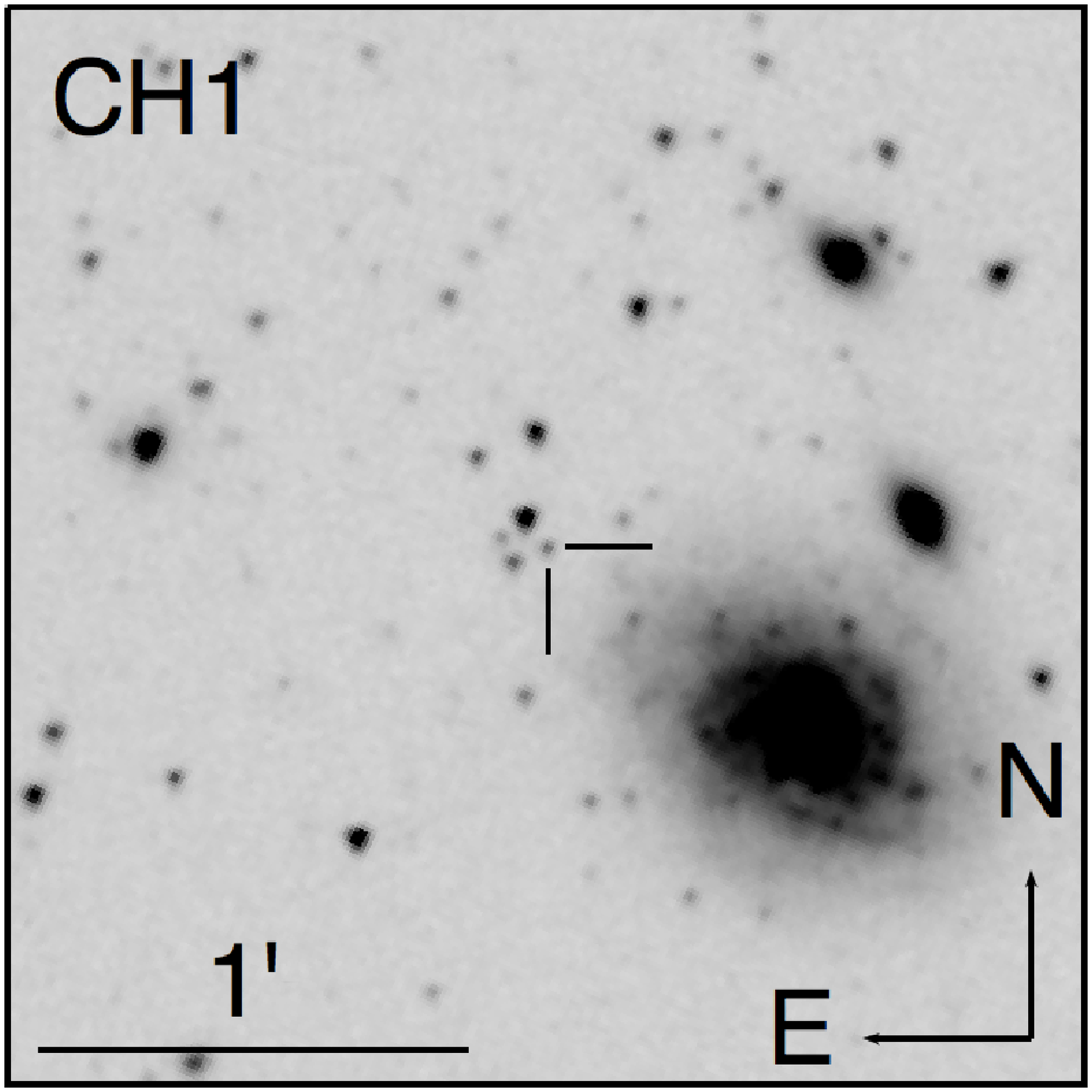}
}
\hspace{1mm}
\subfigure[4.5$\upmu$m]{
\includegraphics[height=41mm,width=41mm]{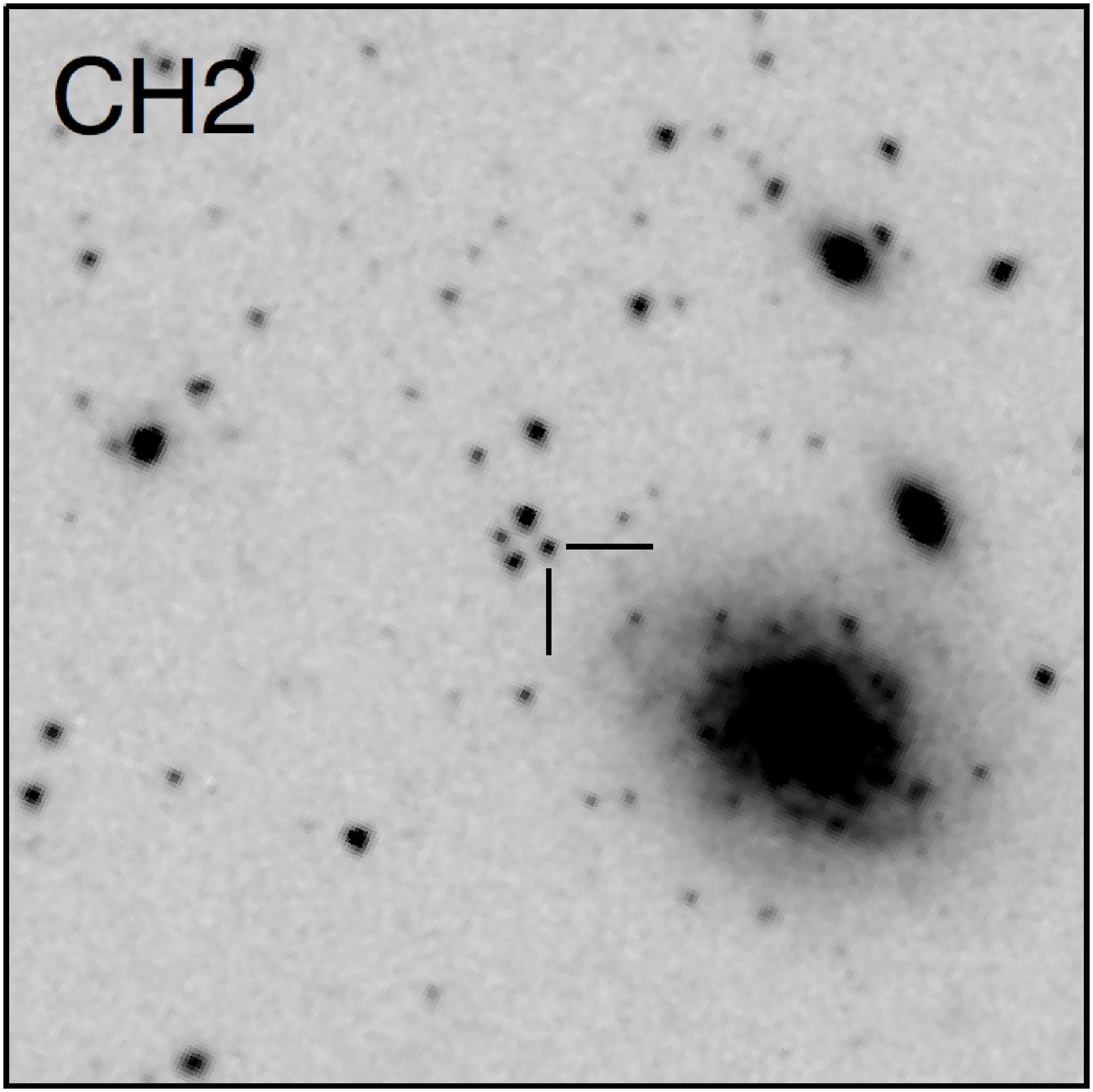}
}
\caption[]{MIR images of SN 2009ip, taken with the {\it Spitzer Space Telescope} + IRAC on 2013 July 25.}
\label{fig:mir}
\end{figure}

\subsection{Spectroscopy}

We obtained a series of intermediate resolution optical spectra of SN 2009ip between 2013 April and December, together with a subsequent set of spectra from the 2014 observing season covering the period 2014 April to September, as listed in Table \ref{tab:spec}. Longslit spectra were taken with the NTT+EFOSC2 (as part of the PESSTO project\footnote{PESSTO data taken up to 2014 April 30 are available from the ESO archive as Phase III products (within the Spectroscopic Survey Data Release 1; SSDR1), and the remainder of the data will be released in 2015 as SSDR2. Detailed instructions on obtaining PESSTO data is available on the PESSTO website, www.pessto.org.}), the 8.2m Very Large Telescope (VLT) + FOcal Reducer and Spectrograph 2 (FORS2), the 8.1m Gemini South telescope + GMOS2, and the 10m Keck II telescope +  the DEep Imaging Multi-Object Spectrograph \citep[DEIMOS;][]{Fab03}. Higher resolution echelle spectra were taken with VLT+ XShooter \citep{Ver11}; the XShooter spectra cover a broad wavelength range from the atmospheric cut off in the blue, to 2.5$\upmu$m. We also obtained three spectra with the ANU 2.3m telescope + Wide-Field Spectrograph (WiFeS), these had low S/N even after rebinning to a lower resolution, but are shown for completeness in Fig. \ref{fig:optspec}.

All EFOSC2 spectra were reduced using the PESSTO pipeline \citep{Sma14}, while the FORS2 and GMOS spectra were reduced within {\sc iraf} using the {\sc longslit} and {\sc gemini gmos} packages respectively. In all cases, the data were reduced in the standard fashion: extracted, wavelength and flux calibrated, and in the case of the EFOSC2 spectra corrected for telluric absorption. The WiFeS spectra were reduced using the {\sc pywifes} pipeline \citep{Chi14}; as WiFeS is an integral field unit spectrograph, one dimensional spectra were extracted from the data cube using a PSF-weighted fit. The XShooter spectra were reduced with the ESOREX recipes run under the {\sc reflex} environment. For all XShooter spectra, a solar analog was observed at comparable airmass to SN 2009ip and was used to correct for the regions of low atmospheric transmission in the NIR. In the final XShooter spectrum obtained of SN 2009ip in 2014 August, no flux from SN 2009ip was detected in the NIR arm.

\begin{figure*}
\centering
\includegraphics[width=0.9\linewidth,angle=0]{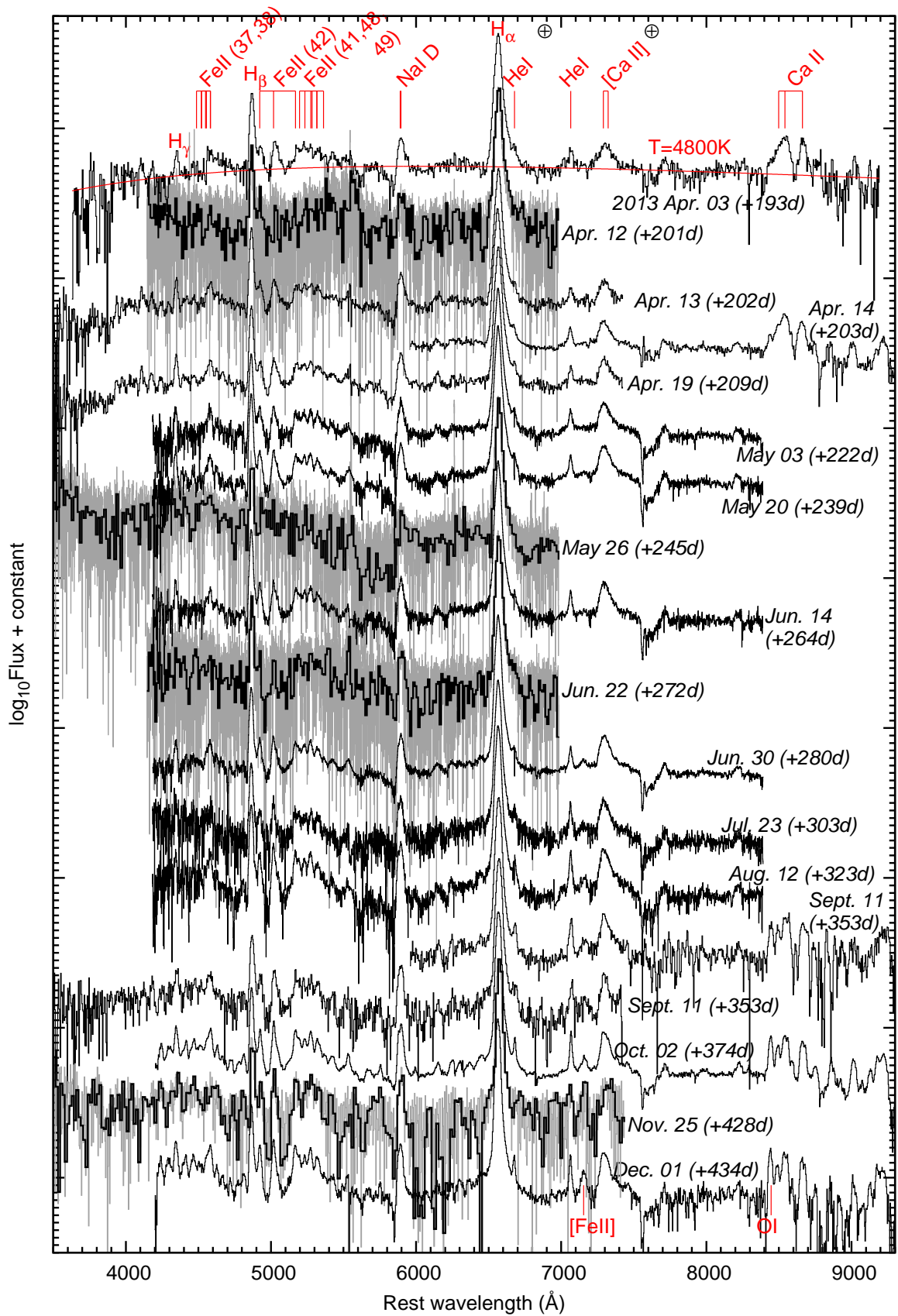}
\caption{Sequence of optical spectra of SN 2009ip taken between 2013 April and December, with phases relative to MJD 56193 in parentheses. All spectra are shown in the rest frame, the wavelengths of key features are marked, while the positions of the strong telluric absorptions are shown with a $\oplus$ symbol. The WiFeS spectra and the EFOSC2 spectrum from November 25 are shown in light grey, with a binned version of the spectrum overplotted in black. A 4800K blackbody fit to the 2013 April 3 spectrum is shown in red. The higher resolution spectra taken during this period are shown in more detail in Fig. \ref{fig:highresspec}.}
\label{fig:optspec}
\end{figure*}

\begin{figure*}
\centering
\includegraphics[width=0.9\linewidth,angle=0]{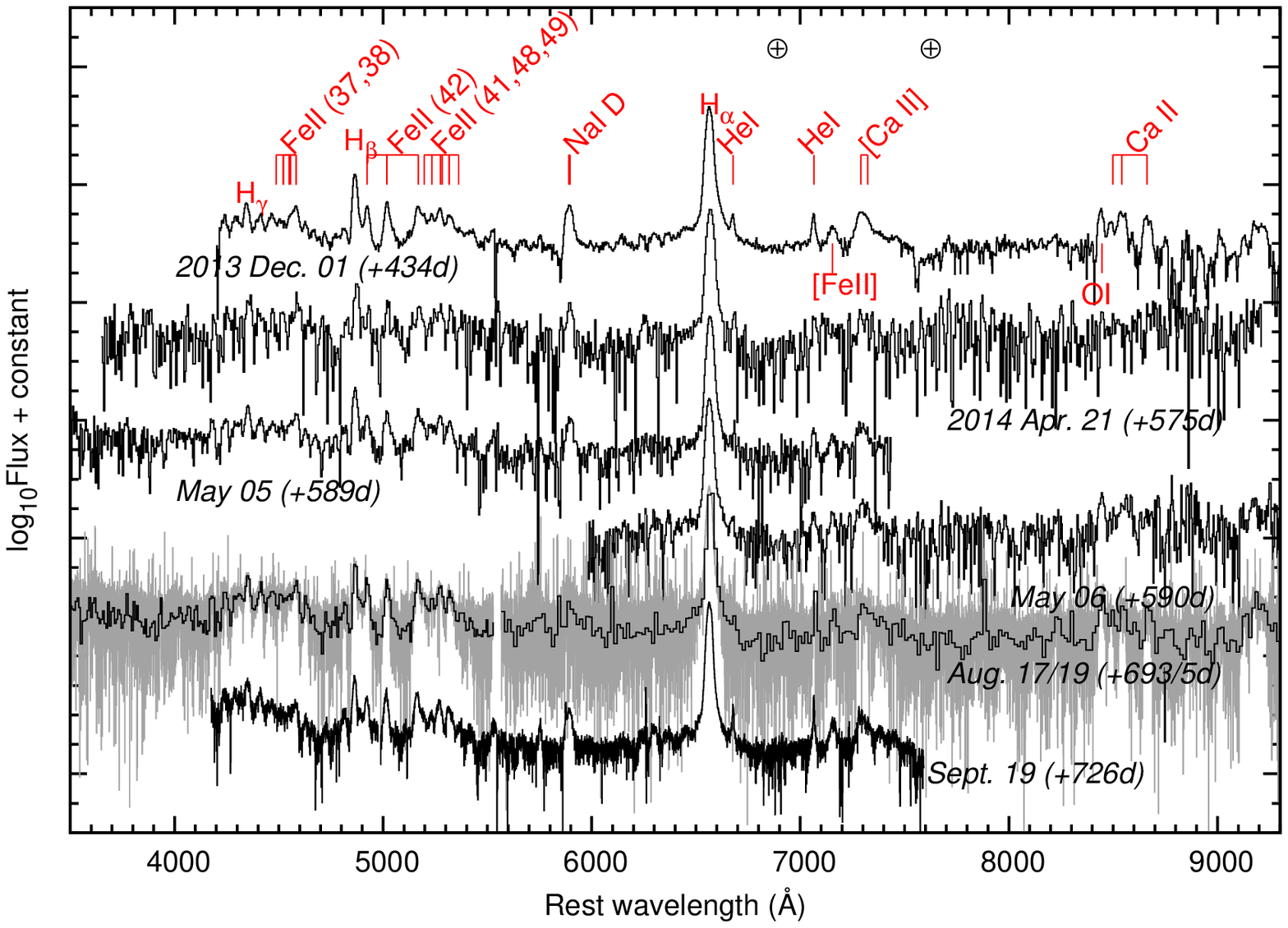}
\caption{Sequence of optical spectra of SN 2009ip taken between 2014 April and September, with phases relative to MJD 56193 in parentheses. All spectra are shown in the rest frame, the wavelengths of key features are marked, while the position of the strong telluric absorptions are shown with a $\oplus$ symbol.  The EFOSC2 spectrum from Aug 17/19 is shown in light grey, with a binned version of the spectrum overplotted in black. For comparison, the last spectrum taken of SN 2009ip in 2013 December (i.e. the final spectrum in Fig. \ref{fig:optspec}) is also shown. The higher resolution spectra taken during this period are shown in more detail in Fig. \ref{fig:highresspec}.}
\label{fig:optspec_2014}
\end{figure*}

\begin{table*}
\caption{Log of spectroscopic observations of SN 2009ip.}
\begin{center}
\begin{tabular}{lllcccr}
\hline
Phase (d)	&Date		& MJD\footnote{midpoint of exposure}			& Telescope	& Instrument	& Grism			& Range (\AA)	\\
\hline
193	& 2013-04-04.39	& 56386.39		& NTT		& EFOSC2	& Gr13			& 3650--9246	\\
201	& 2013-04-12.00	& 56394.00		& ANU 2.3m	& WiFeS		& B3000+R7000	& 4171--7021	\\
202	& 2013-04-13.40	& 56395.40		& NTT		& EFOSC2	& Gr11			& 3346--7466	\\
203	& 2013-04-14.39	& 56396.39		& NTT		& EFOSC2	& Gr16			& 5995--9990	\\
207  & 2013-04-18.39	& 56400.39		& NTT		& SOFI		& BG			& 9277--16345	\\
209	& 2013-04-20.39	& 56402.39		& NTT		& EFOSC2	& Gr11			& 3347--7467	\\
222	& 2013-05-03.40	& 56415.40		& Gemini-S	& GMOS		& R400			& 4208--8439	\\
226	& 2013-05-06.35	& 56418.35		& VLT		& XShooter	& UVB+VIS+NIR	& 3000--24800	\\
239	& 2013-05-20.35	& 56432.35		& Gemini-S	& GMOS		& R400			& 4207--8439	\\
242	& 2013-05-23.43	& 56435.43		& Gemini-S	& GMOS		& B600			& 3191--6076	\\
245	& 2013-05-26.00	& 56438.00 		& ANU 2.3m	& WiFeS		& B3000+R7000	& 3500--7023	\\
264	& 2013-06-14.36	& 56457.36 		& Gemini-S	& GMOS		& R400			& 4208--8437	\\
272	& 2013-06-22.00	& 56465.00		& ANU 2.3m	& WiFeS		& B3000+R7000	& 4169--7023	\\
280	& 2013-06-30.43	& 56473.43		& Gemini-S	& GMOS		& R400			& 4210--8441	\\
303	& 2013-07-23.38	& 56496.38		& Gemini-S	& GMOS		& R400			& 4207--8440	\\
323	& 2013-08-12.19	& 56516.19	 	& Gemini-S	& GMOS		& R400			& 4208--8439	\\
353	& 2013-09-12.05	& 56547.05		& NTT		& EFOSC2	& Gr11+Gr16		& 3343--9986	\\
374	& 2013-10-03.14	& 56568.14		& VLT		& FORS2		& 300V			& 4226--9639	\\
428	& 2013-11-26.08	& 56622.08		& NTT		& EFOSC2	& Gr11			& 3365--7478	\\
434	& 2013-12-02.06	& 56628.06		& VLT		& FORS2		& 300V			& 4224--9632	\\
575	&2014-04-22.37	& 56769.37		& NTT		& EFOSC2	& Gr13			& 3650--9246	\\
589	& 2014-05-06.36	& 56783.36		& NTT		& EFOSC2	& Gr11			& 3365--7478	\\
590	& 2014-05-07.33	& 56784.33		& NTT		& EFOSC2	& Gr16			& 5995--9990	\\
590	& 2014-05-07.39	& 56784.39		& VLT		& XShooter	& UVB+VIS+NIR	& 3000--24800	\\
693/5 & 2014-08-17/19	& 56887.16		& VLT		& XShooter	& UVB+VIS+NIR	& 3000--24800	\\
726   & 2014-09-20.36	& 56920.36 		& Keck		& DEIMOS	& 600ZD			& 4200--7640	\\
\hline	
\end{tabular}
\end{center}
\label{tab:spec}
\end{table*}

NIR spectroscopy was obtained with NTT+SOFI. The SOFI spectrum was taken with the Blue Grism (BG) and reduced using the PESSTO pipeline; the steps involved were flat-fielding, sky-subtraction using pairs of dithered observations and wavelength calibration. Water vapour absorption was identified and corrected for in the spectrum using a telluric standard (a solar analog) which was observed immediately after the spectrum of SN 2009ip, and at a similar airmass. An approximate flux calibration of the spectrum was also made using the telluric standard. We discuss the NIR spectra obtained for SN 2009ip in Section \ref{sect:nirspec}.

\section{Photometric evolution}

The late time lightcurve of SN 2009ip is shown in Fig. \ref{fig:optlc}. SN 2009ip appears to be slowly declining in magnitude, with no further outbursts or re-brightenings over the two years following the 2012a and 2012b events. Extrapolating the lightcurve back to 2012 December, we find that the slow decline is consistent with the last photometry taken at the end of 2012 December. Most notably, the decline rate is significantly slower than the 0.98 mag per 100 days which is typical of the late-time $^{56}$Co-powered tail of a SN. From a linear fit to the {\it R}-band lightcurve in the period +200--450 days (relative to the start of the 2012a event), we measure a decline rate of 0.34 mag per 100 days, while the {\it V}-band yields a decline rate of 0.46 mag per 100 days. During the 2014 observing season, the decline rate is even slower, at 0.27 mag per 100 days in {\it V} and 0.30 mag per 100 days in {\it R}.

The MIR photometry available for SN 2009ip is quite limited, as can be seen in Fig. \ref{fig:optlc}. From 2013 February until 2013 August, the decline in the 3.6 and 4.5$\upmu$m bands appears consistent with that seen in the NIR, and only slightly slower than that seen at optical wavelengths. Over the period 2013 July 25 to August 27, three epochs of MIR photometry were obtained; over this short period we observe a decline at 3.6$\upmu$m, but an apparently constant flux at 4.5$\upmu$m. It is possible that this is caused by contamination of the photometry of SN 2009ip by flux from a nearby source; as can be seen in Fig. \ref{fig:mir}, there are two bright sources close to SN 2009ip which are much brighter in the MIR than at optical wavelengths. Alternatively, if the constant 4.5$\upmu$m flux is due to SN 2009ip, then it may be caused by the formation of CO, which has its fundamental emission in this band.

\begin{figure*}
\includegraphics[width=0.7\linewidth,angle=270]{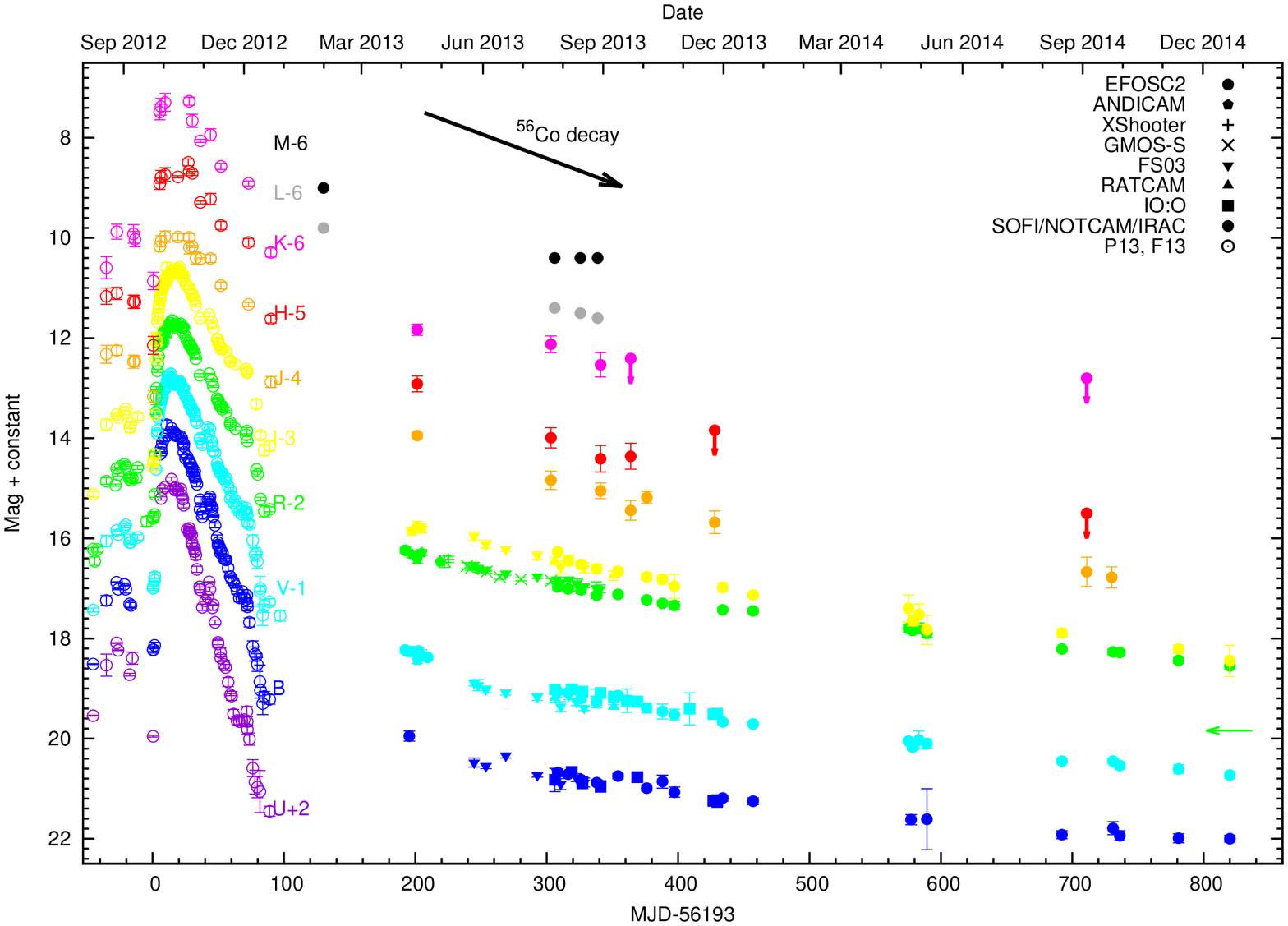}
\caption{Optical and infrared lightcurves for SN 2009ip. The data from \protect\cite{Pas13} and \protect\cite{Fra13} covering the 2012a and 2012b events are shown with open points. The black line shows the typical 0.98 mag per 100 days decline expected during the radioactively powered tail phase of a core-collapse SN. Upper limits where SN 2009ip was not detected are indicated with arrows; the green horizontal arrow at the lower right indicates the presumed quiescent F606W ($\sim${\it R}-band) magnitude measured from archival {\it HST} images by \protect\cite{Fol11}, scaled by $-$2 mag to match the {\it R}-band lightcurve. The source of each point in the optical lightcurve is indicated by the key.}
\label{fig:optlc}
\end{figure*}

The colour of evolution of SN 2009ip is shown in Fig. \ref{fig:col}. We find that in {\it V}$-${\it R}, SN 2009ip has a roughly constant colour over the period from 300 to 800 days after the start of the 2012a event. In contrast, the {\it B}$-${\it V} colour becomes slightly bluer over time. This is likely caused by the presence of emission from Fe group lines in the spectral region covered by the {\it B}-band filter. The {\it V}$-${\it H} colour of SN 2009ip can be measured until $\sim$400 days, after this SN 2009ip is not detected in {\it H}-band, and we can only set upper limits to the NIR colour. The fact that SN 2009ip does not appear to become significantly redder after $\sim$400 days would indicate that dust formation is not significant at this phase.

\begin{figure}
\centering
\includegraphics[width=0.8\linewidth,angle=0]{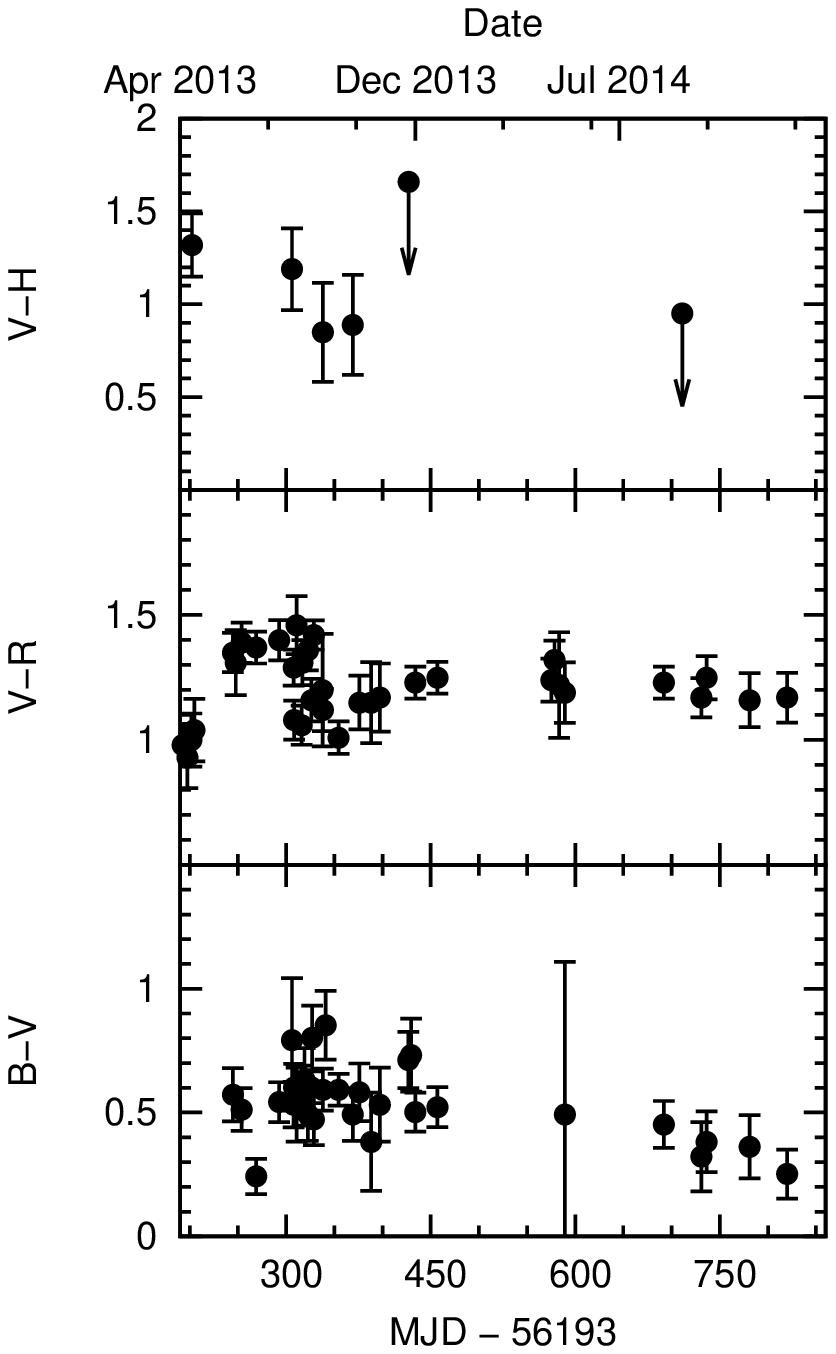}
\caption{The {\it B$-$V}, {\it V$-$R} and {\it V$-$H} colour evolution of SN 2009ip between 2013 April and 2014 December. Limits on the colour of SN 2009ip, from epochs where it was only detected in a single filter, are indicated with arrows. The colour curves have been corrected for foreground extinction.}
\label{fig:col}
\end{figure}

We constructed a pseudo-bolometric {\it BVRIJHKs} lightcurve for SN 2009ip using the same technique as in \cite{Fra13}, which is shown in Fig. \ref{fig:bol}. As expected given the steady decline of the individual photometric bands, the bolometric lightcurve shows a smooth decline. A comparison to the well-studied Type IIP SN 2004et shows that the decline rate of SN 2009ip between $\sim$+250 and +500 days from the start of the 2012a event is slower than typical for a core-collapse SN at a comparable epoch. The degree to which radioactive Co contributes to the energy source of SN 2009ip during this late phase is, however, uncertain. As discussed later, the presence of strong and narrow H$\alpha$ at this epoch suggests that CSM interaction is ongoing, and so one may expect any Co-powered tail to be supplemented by additional flux from the continuing conversion of kinetic energy. The strongest constraint on the Ni mass comes from the lightcurve in late 2012 December;  from a comparison of the bolometric light curves of SNe 2009ip and 2004et at this epoch, we confirm the upper limit to the ejected Ni mass of the former of M$_{\mathrm Ni}<$0.02 \msun\ which was set by \cite{Fra13}.

\begin{figure}
\centering
\includegraphics[width=0.7\linewidth,angle=270]{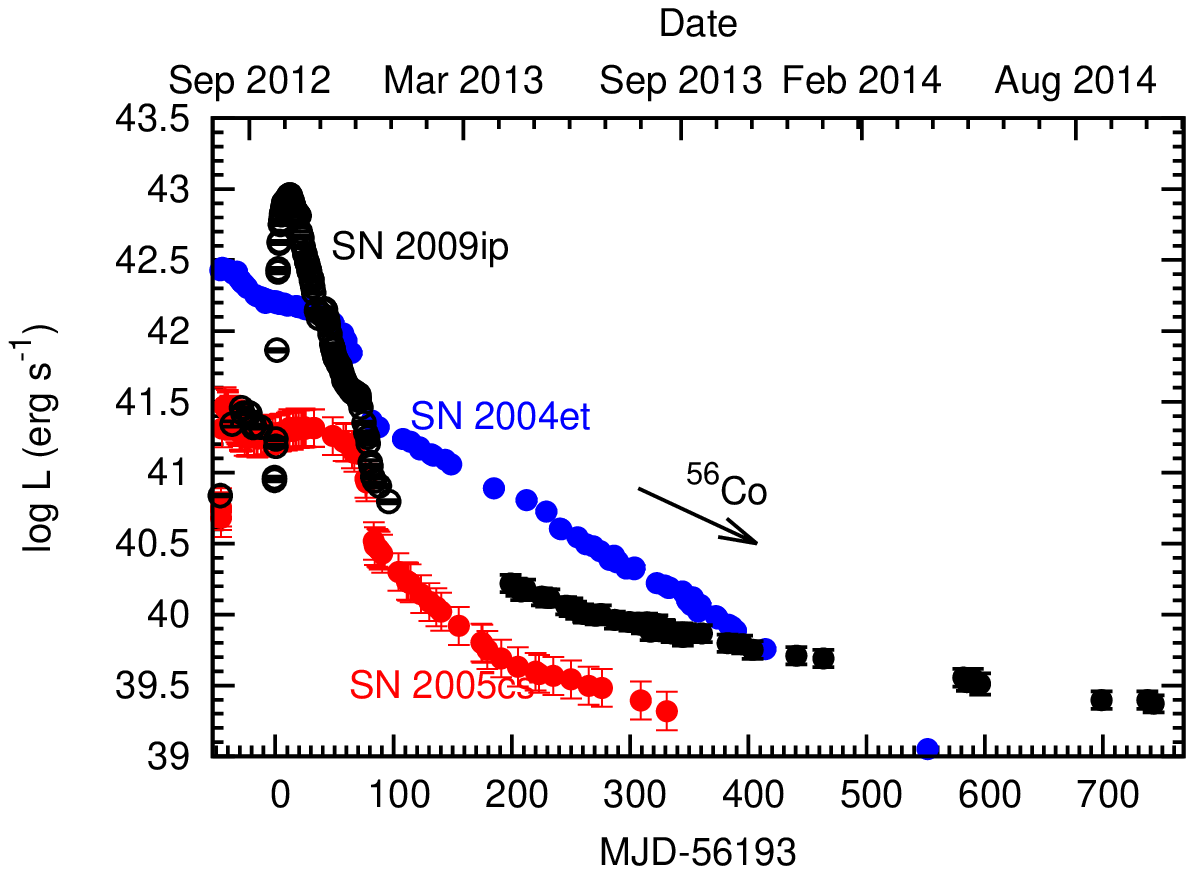}
\caption{{\it BVRIJHK} pseudo-bolometric lightcurve of SN 2009ip \citep[open points;][]{Pas13,Fra13}, plus the additional late-time data presented in this paper (solid points). For comparison we show the core-collapse SNe 2005cs \citep{Pas09} and 2004et \citep{Mag10} shifted so that their explosion epoch matches the start of the 2012a event.}
\label{fig:bol}
\end{figure}

We also calculated bolometric light curves using the optical ({\it BVRI}), NIR ({\it JHK}), and MIR ({\it LM}) bands separately. The {\it Spitzer} 3.6 and 4.5 $\upmu$m fluxes were calibrated to the {\it L} and {\it M} bands using the zeropoints in \cite{Cam85}. For the period where there were MIR observations (2013 January -- August), the contribution at these wavelengths to the total optical-MIR luminosity of SN 2009ip was only $\sim$10 per cent. The NIR contributed a further $\sim$30-35 per cent of the luminosity, while the remainder of the flux emerged at optical wavelengths as shown in Fig. \ref{fig:frac}. We see that after 2013 September, the fraction of flux in the NIR relative to the optical appears to decrease by about 5 per cent. It is possible that this is due to a large amount of the flux in SN 2009ip emerging in line emission, much of which is at optical wavelengths.

\begin{figure}
\centering
\includegraphics[width=0.7\linewidth,angle=270]{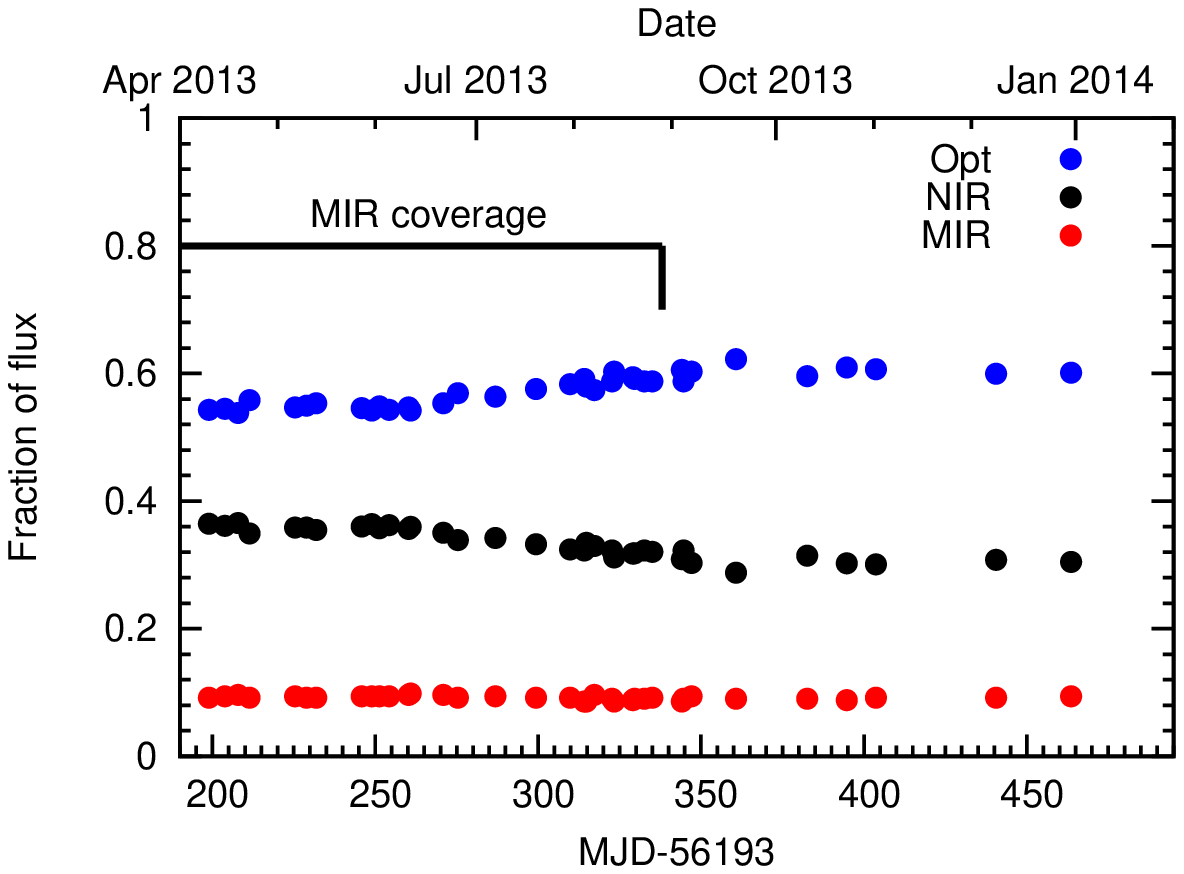}
\caption{The fraction of flux in optical ({\it BVRI}), NIR ({\it JHK}) and MIR ({\it Spitzer}) bands for SN 2009ip during the tail phase. Epochs later than 2013 December have scant NIR coverage (SN 2009ip is detected in {\it J}-band only) and hence are not shown; the time period covered by {\it Spitzer} MIR observations is also indicated.}
\label{fig:frac}
\end{figure}

\cite{Mar15} and \cite{Mar14} put forward the intriguing suggestion that there was a periodic component of the lightcurve of SN 2009ip. With a view to testing this in the late time data, we took the {\it R}-band lightcurve of SN 2009ip between 2013 April 3 and December 24, and made a least-squares linear fit to the data. To remove the overall decline from the lightcurve, a ``residual lightcurve'' was constructed, consisting of the measured magnitude at each epoch minus the interpolated magnitude from the linear fit. The {\it R}-band data were used for this as this was the filter which had the most data points (36 epochs). We then used the Lomb-Scargle algorithm \citep{Lom76,Sca82} to derive a periodogram from the residual lightcurve. The resulting periodogram is shown in Fig. \ref{fig:period}, but as expected from visual inspection of the lightcurve, we see no significant peaks in the power spectrum that would suggest a periodic signal. The absence of any periodic component to the {\it R}-band light curve may indicate that the region of CSM which the ejecta is now interacting with has a different density profile, or was formed through a different mass loss process to that which formed the CSM closer to SN 2009ip. 
The progenitor candidate identified by \cite{Fol11} for SN 2009ip had a magnitude of 21.84$\pm$0.17 mag in {\it F606W}-band images taken in 1999 June. On 2014 December 23, the measured magnitudes of SN 2009ip were {\it V} = 21.73 $\pm$ 0.17 and {\it R}=20.55 $\pm$ 0.07 mag. In {\it V}-band, SN 2009ip reached a similar magnitude to the purported progenitor in late 2014. However, as the {\it HST} images used by \citeauthor{Fol11} were taken with the {\it F606W} filter, they include the wavelength of H$\alpha$. As the H$\alpha$ emission seen in SN 2009ip is strong, even during quiescent periods \citep{Pas13}, it is likely that H$\alpha$ emission is contributing to the flux here also. Hence the most appropriate filter to make a comparison to the pre-discovery {\it HST} images is {\it R}-band, which similarly includes H$\alpha$. SN 2009ip is still $\sim$1.3 mag brighter than the progenitor candidate in {\it R}-band, and assuming a constant decline rate of $\sim$0.3 mag per 100 days, will only reach a comparable magnitude around 2016 March.

\begin{figure}
\centering
\includegraphics[width=1.4\linewidth,angle=270]{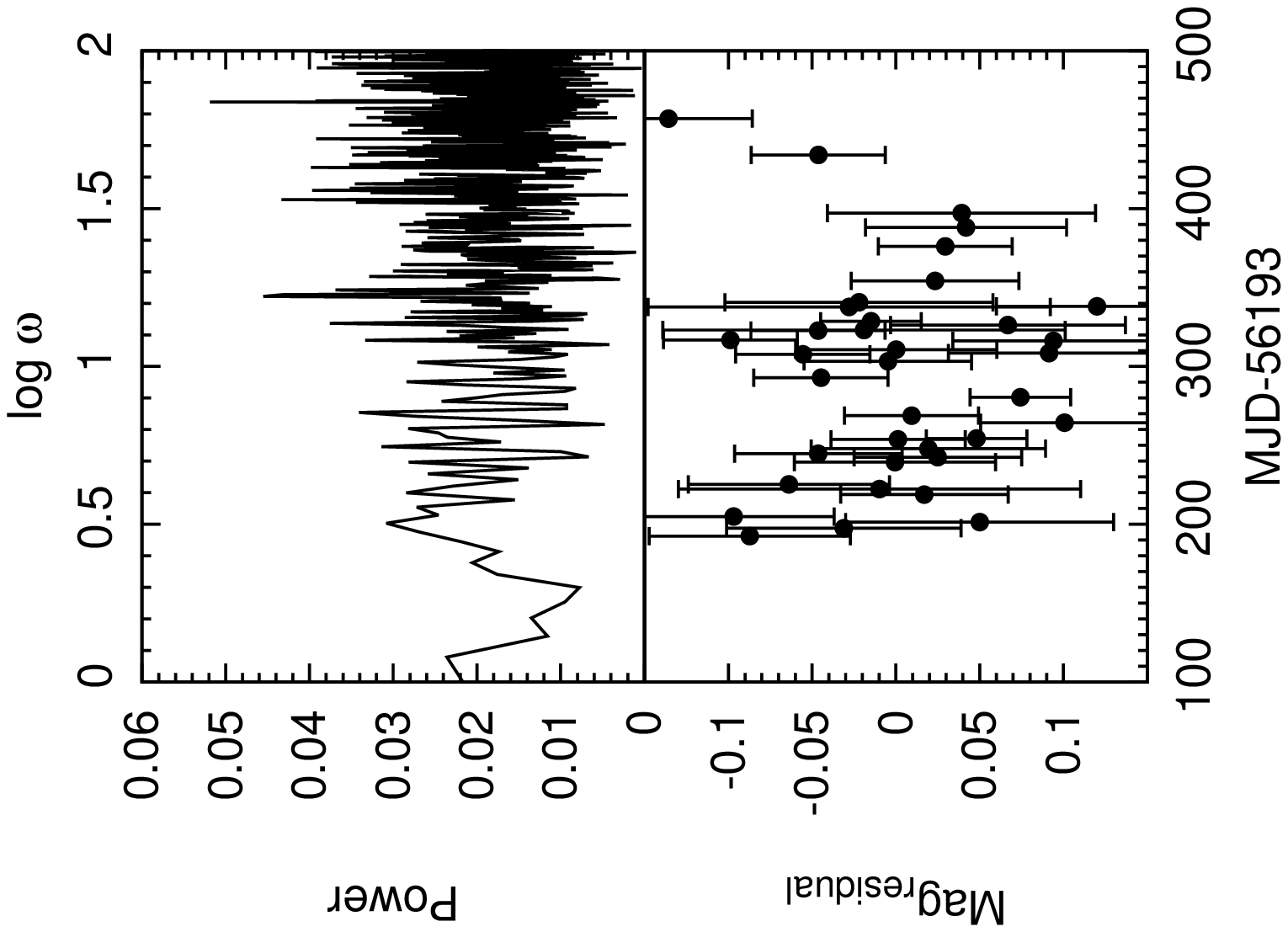}
\caption{Lower panel: {\it R}-band lightcurve of SN 2009ip between 2013 April and December, after the subtraction of a linear least-squares fit to the data points. Upper panel: Lomb-Scargle periodogram of the same, where $\omega$ is the angular frequency in radians per day, and power is in arbitrary units.}
\label{fig:period}
\end{figure}

\section{Spectroscopic evolution}

\subsection{Optical spectra}
\label{sect:specta}

As is evident from Figs. \ref{fig:optspec} and \ref{fig:optspec_2014}, SN 2009ip does not undergo any dramatic spectral evolution between 2013 April and 2014 September. From the first spectrum of 2013, which was taken on April 3, until June 22, we do not observe the emergence of any new lines or significant changes in velocity or line profile shape. In the following, we describe the 2013 April 3 spectrum as representative of SN 2009ip over the subsequent three months. We note that the five highest resolution and signal-to-noise spectra are shown in more detail in Fig. \ref{fig:highresspec}. All of the reduced spectra presented in this paper will be placed in the the WiseRep archive\footnote{http://www.weizmann.ac.il/astrophysics/wiserep/} \citep{Yar12}.

\begin{figure*}
\centering
\includegraphics[width=0.8\linewidth,angle=0]{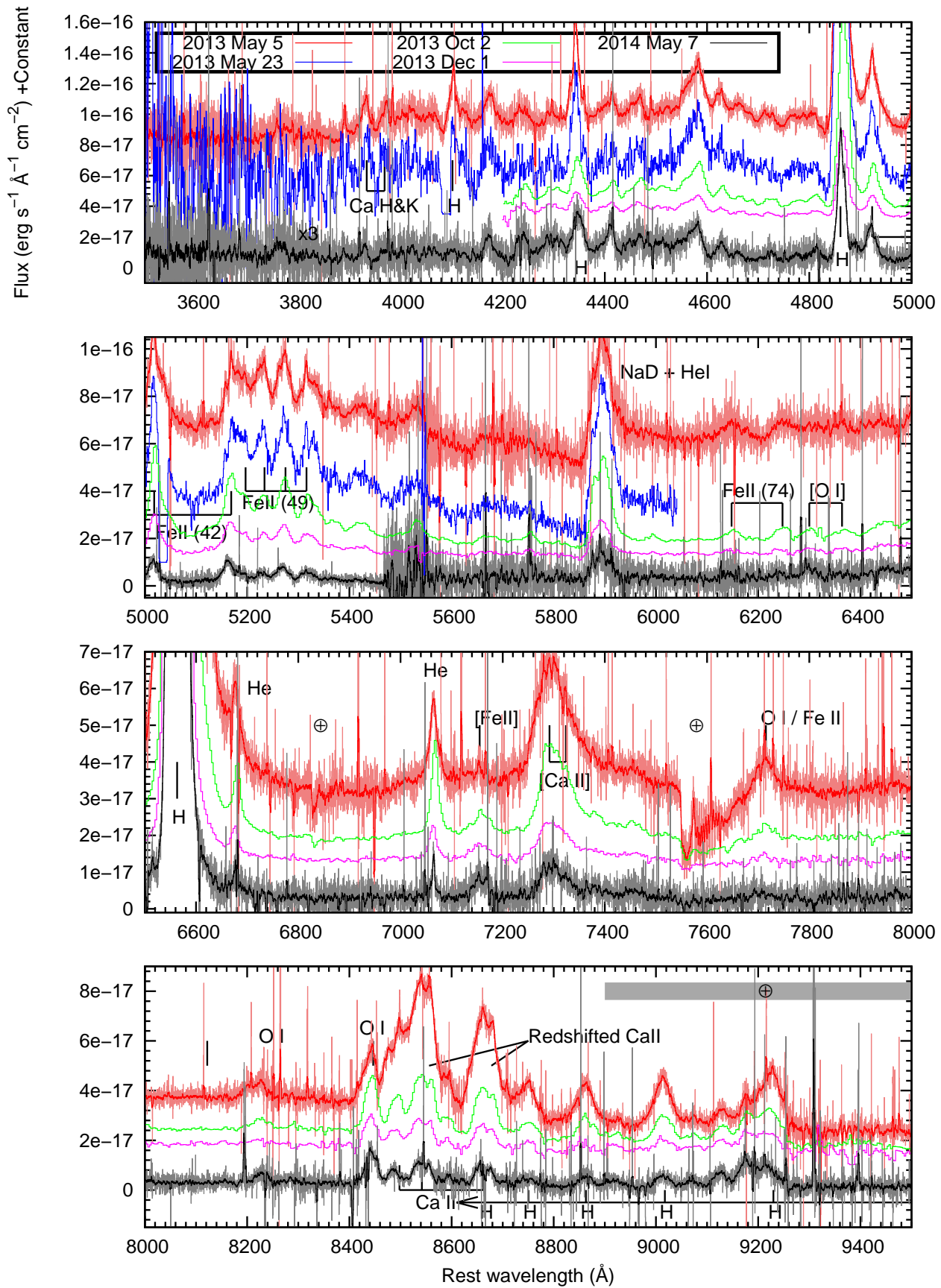}
\caption{Rest frame high resolution VLT+XShooter, VLT+FORS and Gemini+GMOS spectra of SN 2009ip, taken between 2013 May and 2014 May. In the top panel, the VLT+XShooter spectrum from 2014 May 7 has been multiplied by 3 for clarity, all other spectra have been offset in flux as necessary. The 2013 May 6 and 2014 May 7 XShooter spectra have also been boxcar smoothed using a 15 pixel window (red and black lines) to emphasise the weaker features. The telluric A and B bands, and the region of atmospheric absorption between 8900 and 9900~\AA\ are marked with a $\oplus$ symbol.}
\label{fig:highresspec}
\end{figure*}

As shown in Fig. \ref{fig:optspec}, the spectrum from April 3 has a relatively flat continuum. The most prominent feature in the spectrum is H$\alpha$ emission, which is well fitted with a Lorentzian profile of FWHM 1450$\pm$100 \kms. H$\alpha$ appears to be slightly asymmetric, as shown in Fig. \ref{fig:halpha}, with a deficit of flux in the blue wing. We also see H$\beta$ with a similar velocity and line profile, and a weak H$\gamma$ line. While the line profile of H$\alpha$ appears to narrow slowly over the course of our monitoring campaign, for the first three months this evolution is hard to discern. The Balmer decrement ($L_{H\alpha}/L_{H\beta}$) can be used to constrain the physical environment of SN 2009ip \citep{Lev14,Fra13}, and in the April spectrum of SN 2009ip we measure a Balmer decrement of $\sim$12. This is much higher than in 2012, and could indicate Case C recombination, where H$\alpha$ is optically thick \citep{Xu92}. However, the lack of any clear absorption in the Balmer lines would disfavour this scenario. Alternatively, if H$\alpha$ is optically thin, a lower electron temperature and a higher density than at peak could perhaps give rise to the observed decrement \citep{Dra80}\footnote{In \cite{Fra13}, we did not use the results of \citeauthor{Dra80}, as the high luminosity of the 2012b event requires the inclusion of the radiation field when computing line fluxes. Here, as SN 2009ip is much fainter, the assumption of \citeauthor{Dra80} that the radiation field is weak is more appropriate.}. The higher density could be explained as the photosphere recedes inwards through material ejected in a wind or mass loss with a $\propto R^{-2}$ dependence, although it is unclear whether such a simple scenario would necessarily be appropriate to SN 2009ip. The strong H$\alpha$ line is also interesting in the context of the low photospheric temperature. As the spectral energy distribution of SN 2009ip shifts to cooler temperatures than at peak, there are no longer the UV photons necessary to produce recombination lines. It is likely that X-rays from shocks in the interaction region are now supplying the necessary energy to excite H$\alpha$, and perhaps also help maintain the relatively constant blue colour of SN 2009ip (Fig. \ref{fig:col}). In the red part of the spectrum, we identify some of the higher order Paschen lines. These can be seen more clearly in the spectrum from April 13, where Pa lines as high as the 12$\rightarrow$3 transition at $\lambda$8750 can be seen close to the Ca~{\sc ii} NIR triplet.

To estimate the temperature of SN 2009ip in 2013 April, we fitted a blackbody function to the spectra obtained in 2013 April, with the exception of the WiFeS data. The regions of the spectrum around H$\alpha$ and H$\beta$ were excluded from the fit in all cases. We measure an effective temperature of 4800 K at this epoch ($\sim$+200 d). Subsequent spectra show a slight increase in flux to the blue of $\sim$5500 \AA, but this is likely caused by a relative increase in the strength of the Fe~{\sc ii} multiplets between 5000 and 5500 \AA. We observe a similar effect in the  {\it B}$-${\it V} colour. At later epochs, a clear continuum arising from a photosphere cannot be discerned, and so we cannot measure a physically meaningful effective temperature.

\begin{figure}
\centering
\includegraphics[width=0.9\linewidth,angle=270]{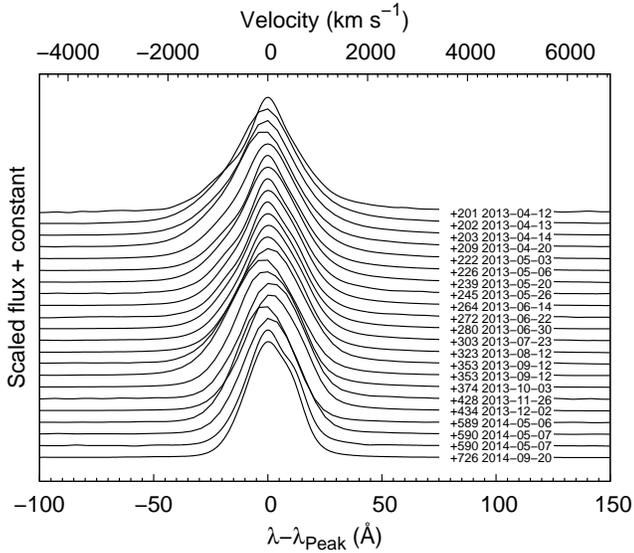}
\caption{Evolution of the H$\alpha$ line profile over time. The sequence contains all optical spectra from Table \ref{tab:spec} which cover the wavelength of H$\alpha$, with the exception of the EFOSC2 Gr\#13 spectra which were of lower resolution. The date of each spectrum is indicated, as is the phase relative to the start of the 2012b event, MJD=56193. All spectra have been convolved with a Gaussian to degrade them to a common resolution of R=1000, have been normalised to the peak of H$\alpha$, and have been shifted so that the peak of H$\alpha$ is at 0~\AA. The emergence of a red shoulder in H$\alpha$ at at the end of 2014 is apparent in the final two spectra.}
\label{fig:halpha}
\end{figure}

\begin{table}
\caption{Measurements of the FWHM of H$\alpha$ in SN 2009ip, as found from a Gaussian fit to the emission.}
\begin{center}
\begin{tabular}{llc}
\hline
Phase (d)	&Date		& FWHM (\AA)	\\
\hline
202	&   2013-04-12	&  34.3	\\
201	&   2013-04-13	&  37.1	\\
203	&   2013-04-14	&  36.3	\\
209	&   2013-04-20	&  36.7	\\
222  &   2013-05-03  &  35.1	\\
226	&   2013-05-06	&   33.1	  \\
239	&   2013-05-20	&  34.4	\\
245	&   2013-05-26	&  32.7	\\
264	&   2013-06-14	&  33.2	\\
272	&   2013-06-22	&   32.3	 \\
280	&   2013-06-30	&  33.0	\\
303	&   2013-07-23	&  33.0	\\
323	&   2013-08-12	&  32.7	\\
374	&   2013-10-03	&  32.2	 \\
428	&   2013-11-26	&  31.2	\\
434	&   2013-12-02	&  30.1	\\
589	&   2014-05-06	&  28.2	\\
590	&   2014-05-07	&  28.7	\\
590	&   2014-05-07	&  26.2	\\
726	&   2014-09-20	&  25.3	\\
\hline
\end{tabular}
\end{center}
\label{tab:fwhm}
\end{table}

In addition to H, we see He, Na, Ca, and Fe in the spectrum of SN 2009ip in 2013 April. Weak He~{\sc i} $\lambda$7065 is also present in emission. The other transitions of He in the optical are unfortunately blended with other lines; He~{\sc i} $\lambda$5876 is blended with Na~{\sc i}~D, while both the $\lambda$4922 and $\lambda$5015 He~{\sc i} lines lie in the region of strong emission from Multiplet 42 of Fe. We also tentatively detect He~{\sc i} $\lambda$6678 which was seen at earlier epochs \citep{Fra13}. The He~{\sc i} $\lambda$7065 line is not particularly clear in the April 3 spectrum, and so we instead measure its FWHM in the April 14 spectrum taken with EFOSC2 + Gr\#16, where we find it to have a FWHM of only $\sim$1000 \kms, which is significantly slower than H$\alpha$. The O~{\sc i}  $\lambda$8446 line can also be seen to emerge in the April 13 spectrum, with a FWHM of $\sim$1250~\kms. This feature was possibly present in the earliest spectrum taken on April 3, albeit blended with the blue wing of Ca~{\sc ii} $\lambda$8498. There is perhaps some contribution to the line from the 18$\rightarrow$3 transition of H ($\lambda$8438), although the intensity of the feature in comparison to the other Pa lines suggests that H is not the dominant element contributing to any blend. Furthermore, the feature grows in strength over the following six months, consistent with the line being formed primarily by O.  O~{\sc i} $\lambda$8446 emission in Type II SNe typically arises via Lyman~$\beta$ pumping in O~{\sc i} $\lambda$1026 \citep[e.g.][]{Oli93}. The strength of this line in Type II SNe has been interpreted as radiative cross-talk between H and O zones \citep{Oli93,Jer12}, although unless it is very strong, the $\lambda$8446 line (and O~{\sc i} 1.129$\upmu$m)  can be produced Ly-beta scattering in primordial material.

The only line with a clear absorption component is Na~{\sc i}~D, which has a P Cygni profile with a minimum at $-$3000 \kms. The peak of the emission is at 5896~\AA, indicating that the emission is definitely caused by Na, although we caution that the absorption may be contaminated by He~{\sc i} $\lambda$5876. There is also a tenuous detection of a second absorption component in Na at  $\sim$15000 \kms, which would match to the two component absorption seen in earlier epochs \citep{Fra13}. Na~{\sc i}~D tends to remain optically thick for a long time, as it is a resonance line from the ground state, and so it is unsurprising that we continue to see absorption. [Ca {\sc ii}] $\lambda\lambda$7291,7324 is present, as is the Ca~{\sc ii} NIR triplet in pure emission. [Ca~{\sc ii}] appears broader than other features, which is due to the blending of the two components in a doublet, and possibly due to some contamination from He~{\sc i} $\lambda$7281. We measure a FWHM of 1700 \kms\ for the $\lambda$8662 line in the Ca~{\sc ii} NIR triplet, similar to the H and He velocities. Aside from these lines, the only other features present are from Fe, where we see several permitted multiplets in Fe~{\sc ii}, the strongest of which is Multiplet 42 \citep{Moo45}; although there is undoubtedly contamination from He~{\sc i} $\lambda$4922 and $\lambda$5015. We also see a possible weak feature at $\sim$7720~\AA, although this region is affected by telluric absorption. The feature is offset by $-$2000 \kms\ from the rest wavelength of O~{\sc i} $\lambda$7774, and an alternative identification could be Fe~{\sc i} $\lambda$7710. The permitted Fe and Na, and the permitted and forbidden Ca can all likely be accounted for by primordial metals in the envelope of the star \citep{Li93,Mag12,Jer12}\footnote{Although \cite{Mag12} and \cite{Jer12} presented models for the lower mass red supergiant progenitor of a Type IIP SN, the conclusion that much of the line emission seen at late times arises from the envelope should also apply here.}; we stress that these elements were all seen at comparable velocities by \cite{Pas13} in spectra of SN 2009ip taken two years prior to the 2012a and 2012b events, and hence cannot be unambiguously associated with the products of SN nucleosynthesis. The similar velocities measured for H, He and Ca are also consistent with these lines forming in the same location. We also note that whatever the identification of the $\sim$7720~\AA\ feature, both Fe~{\sc i} $\lambda$7710 and O~{\sc i} $\lambda$7774 were seen in the September 2009 spectra \citep{Pas13}. The only ambiguous feature in the spectra of SN 2009ip in 2013 April is a line around $\sim$4580~\AA, which could possibly be identified with Mg~{\sc i}] $\lambda$4571. The feature is centered 600 \kms\ to the red of the rest wavelength of Mg. As discussed later, we do not see the Mg~{\sc i} 1.5 $\upmu$m line in the NIR spectra taken around this epoch, although the NIR line is a recombination rather than a cooling line.

In the XShooter and GMOS spectra obtained in 2013 May, we observe the first evolution in the spectra of SN 2009ip, with the strengthening of the  $\lambda$8446 O~{\sc i} line (Fig. \ref{fig:highresspec}), which was only marginally detected in 2013 April. By 2013 September, the line has grown in strength to be comparable with the components of the close by Ca~{\sc ii} NIR triplet. In the June 30 spectrum, we see the emergence of [Fe~{\sc ii}] $\lambda$7155. The feature is quite weak, but continues to grow in strength until our latest spectrum in December. Interestingly, the $\lambda$7155 feature was not seen in the spectra covering the pre-2012 evolution of SN 2009ip \citep{Pas13}, while the line is often seen during the nebular phase of core-collapse SNe.

The final significant change in the spectrum of SN 2009ip in late 2013 is the emergence of what appears to be weak [O~{\sc i}] $\lambda\lambda$6300,6364 in the VLT+FORS spectrum taken on October 2. Emission lines consistent with both components of the doublet are seen, an appear to have approximately the same intensity. These lines were also seen by \cite{Fox15} in their spectrum of SN 2009ip from 2014 June. The lines are quite weak but clearly real; they are detected in two separate spectra, most clearly on October 2, where the spectrum has a S/N in the continuum of $\sim$25. The lines are also comparable in strength to the nearby Multiplet 74 of Fe~{\sc ii}, as can be seen in Fig. \ref{fig:OII}. As an alternative identification, the presumed O lines could be associated with Fe~{\sc ii}, however we were unable to find a consistent match to any of the common multiplets \citep{Moo45}. Multiplet 40 has a line at 6369~\AA\ which is a reasonable match, however there should also be stronger emission from the same multiplet at $\lambda$6432 (which is possibly blended with the line at 6445 \AA), and at $\lambda$6516 (which may be lost in the blue wing of H$\alpha$. Multiplet 200 has a component at $\lambda$6305, but again this should be accompanied by a line of approximately the same strength at $\lambda$6175 which we do not observe. In the October 2 spectrum, we fitted Lorentzian profiles to the two components of the O doublet and measured an average FWHM of 1400 \kms\, while in the December 1 spectrum we measured 1200 \kms\ and 1400 \kms\ for the $\lambda$6300 and $\lambda$6364 lines respectively. This velocity is similar to that seen in the nearby Fe Multiplet 74 at 1400 \kms, and consistent with that of H$\alpha$.

A ratio of 1:1 between the intensity of the two components of [O~{\sc i}] $\lambda\lambda$6300,6364 would suggest optically thick conditions where the lines are forming; if the lines are forming in the ejecta, then the line ratio would be expected to increase to 3:1 over time as the ejecta expands and the density decreases \citep[e.g][]{Chu92}. Primordial [O~{\sc i}] $\lambda\lambda$6300,6364 is always observed in the 3:1 ratio, and hence an observed ratio of $\sim$1:1 is suggestive of synthesized oxygen which has a higher density. However, as can be seen in Fig. \ref{fig:OII}, its also quite possible that there is contamination of the $\lambda$6364 line by the many iron-group lines in this region. Furthermore, in core-collapse SN spectra the [O~{\sc i}] lines are much stronger than these surrounding Fe lines; and so if this is synthesized oxygen its mass must be very small. In later subsequent spectra taken in 2014, these lines are not detected, as can be seen in Fig. \ref{fig:OII}.

The XShooter and GMOS spectra obtained in 2013 October also reveal some structure in the Ca lines. As can be seen in Fig. \ref{fig:highresspec}, all three lines of the Ca~{\sc II} NIR triplet show an additional emission component, offset by $\sim$550 \kms\ to the red. This feature is real, as it can be seen in multiple spectra at different epochs and from different instruments. It is unclear whether the redshifted components are also present in the [Ca {\sc ii}] $\lambda\lambda$7291,7324 lines; as these lines are only separated by $\sim$30~\AA, they appear blended together and any additional component would be hard to disentangle. A similar redshifted emission component is not seen in any of the other elements, with H, He and Fe all showing relatively symmetric profiles. The lines of O are weak, while Na~{\sc i}~D is blended with He, making it difficult to assess their profiles, however neither of these lines shows an obvious emission shoulder either.

\begin{figure}
\centering
\includegraphics[width=1\linewidth,angle=0]{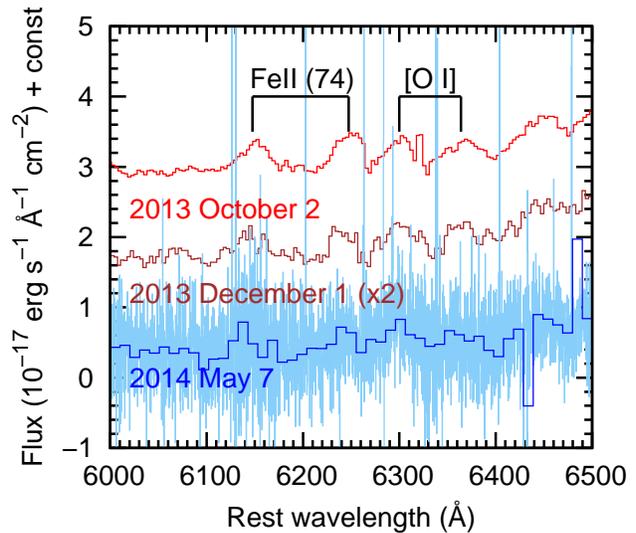}
\caption{The region of the presumed [O~{\sc i}] $\lambda\lambda$6300,6364 doublet in two late time spectra of SN 2009ip, taken in 2013 October and December. The rest wavelengths of the two components of the doublet, along with those of the nearby Fe~{\sc ii} Multiplet 74 are indicated. The 2014 May 7 spectrum is also shown in light blue, but even after binning (dark blue line) the [O~{\sc i}] doublet is not visible at this epoch.}
\label{fig:OII}
\end{figure}

SN 2009ip was behind the Sun during the period from 2014 January to March. When SN 2009ip was first observed again in 2014 April, it appeared to be relatively unchanged from the previous spectrum in 2013 December (Fig. \ref{fig:optspec_2014}). Interestingly, and contrary to na\"ive expectations if SN 2009ip had undergone core-collapse, the [O~{\sc i}] $\lambda\lambda$6300,6364 feature had not strengthened since its first appearance in 2013 October. In fact, only the $\lambda$6300 component of the doublet was detected in the XShooter spectrum from 2014 May. Whether the na\"ive expectation that [O~{\sc i}]  would continue to strengthen is, of course, uncertain; the models of \cite{Jer14} only cover the period up to $\sim$450 days after core-collapse, by which point the [O~{\sc i}] lines appear to have reached their maximum strength.

The subsequent spectrum we obtained of SN 2009ip using VLT+XShooter in 2014 May also revealed an additional emission component in the red wing of the Balmer lines. The feature is seen to be offset by +500 \kms\ from the rest wavelength of H$\alpha$, and is also seen as a faint shoulder in H$\beta$. \cite{Gra14} also saw a slight excess of flux in the red wing of H$\alpha$ in as early as 2013 June, although at this stage the asymmetry was less clear than the distinct shoulder seen in our data from 2014 May. The velocity of the red emission feature in H$\alpha$ is similar to that seen in the second component of the Ca~{\sc II} NIR triplet in 2014 October; we note that the redshifted Ca emission is also still present at this epoch. We also note that the He emission lines in the 2014 May XShooter spectrum are extremely narrow, with He~{\sc i} $\lambda$7065 having a velocity of only $\sim$180 \kms\ as measured from the FWHM of a Lorentzian fit to the line. This is marked contrast to the velocity of 750 \kms\ measured in the same line in the XShooter spectrum taken a year earlier in 2014 May. It is also significantly lower than the velocities seen in the Balmer lines, which have FWHM of around 1000 \kms\ in 2014. We also see similarly low velocities in the line at 5016~\AA\, suggesting that this is dominated by He~{\sc i} $\lambda$5016 at this stage, rather than by Fe~{\sc ii} Multiplet 42.

Our final spectrum of SN 2009ip was obtained on 2014 September 19, and is similar to those from 2014 April and May. The spectrum has an excess of flux in the blue, which we attribute to a pseudo-continuum formed of many weak Fe-group lines. H$\alpha$ and H$\beta$ are still present, with narrow ($\sim$1000 \kms) Lorentzian profiles. [Ca~{\sc ii}] is present, while it is unclear whether Na is still detected due to the He emission at 5876~\AA\ which could mask the presence of Na~{\sc i} D. Finally, the He lines are extremely narrow, with velocities of only $\sim$200 \kms. The lower velocities for the He lines when compared to H suggest that these may arise from different regions. Such relative strength of the He lines may also point to a CSM which is richer in He than a typical Type IIn SN.

We measured the FWHM and central wavelength of H$\alpha$ in all the spectra obtained of SN 2009ip; these measurements are listed in Table \ref{tab:fwhm}. Each spectrum was convolved to a common resolution, and a Gaussian was fit to H$\alpha$, allowing the central wavelength, FWHM and a normalisation constant to vary as free parameters. We see no systematic shift in the position of H$\alpha$ over time, suggesting that substantial dust formation is not occurring. The measured values for the FWHM of H$\alpha$ are plotted in Fig. \ref{fig:velocity}. From Fig. \ref{fig:velocity}, the decrease in FWHM of H$\alpha$ over time, from $\sim$1600 \kms\ in 2013 April, to $\sim$1100 \kms\ in 2014 September, is clear.

While \cite{Smi13} suggested that dust formed in the material lost by SN 2009ip during previous eruptions, we find no evidence for dust formation since the peak of the 2012b event. In the spectra obtained for SN 2009ip between 2013 April and 2014 September, H$\alpha$ appears to be centred on its rest wavelength, with no systematic blueshift which would be indicative of dust. Furthermore, we see no sign of a deficit of flux in the red wing of H$\alpha$. We also do not observe any shifts or deficit in the red wing of any of the other lines. The H-K colour is positive, while J-H is negative, and the V-K colour is only 1.5, providing ambiguous evidence for a K-band excess. Furthermore, as the flux in the continuum of SN 2009ip is relatively weak compared to the strong line emission, it is possible that broad band photometry does not provide a good estimate of the shape of the underlying SED.

\begin{figure}
\centering
\includegraphics[width=1\linewidth,angle=0]{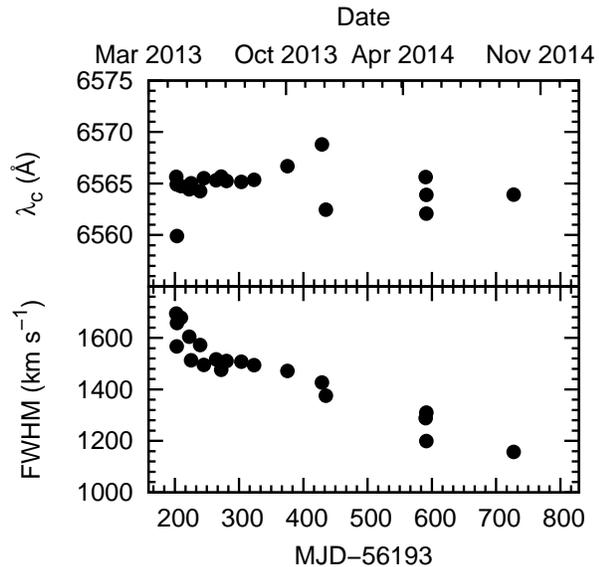}
\caption{The central wavelength and FWHM of H$\alpha$. These values were measured for all spectra via a Gaussian fit to the line.
}
\label{fig:velocity}
\end{figure}

\subsection{NIR spectra}
\label{sect:nirspec}

The NIR spectra of SN 2009ip are shown in Fig. \ref{fig:nir}, while in Fig. \ref{fig:co} we show a zoomed in region of the 2013 May XShooter spectrum, covering the location of the first overtone of CO. The spectrum taken with the NIR arm of XShooter in 2014 May suffers from extremely low S/N, and hence is not discussed here. The two spectra from 2013 were taken less than a month apart (with no apparent evolution between them), and are dominated by H, and to a lesser extent He. We see the Pa lines, including Pa$\alpha$ which is of sufficient strength to be recovered even with the strong high atmospheric absorption between 1.8 and 2.0 $\upmu$m. Lines from the Brackett series are visible in the XShooter spectrum. We also observe the two strongest He~{\sc i} lines at 1.08 and 2.05 $\upmu$m.

\begin{figure*}
\includegraphics[width=0.7\textwidth]{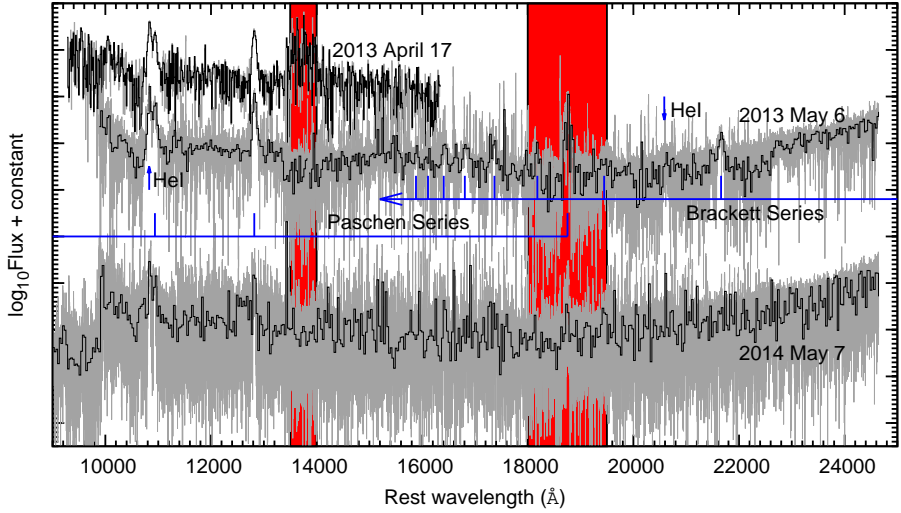}
\caption{NIR spectra of SN 2009ip. To aid in the identification of lines, a binned version of each spectrum is shown in black, while the unbinned spectra are in grey. The red shaded areas correspond to low atmospheric transmission. The final XShooter spectrum taken in 2014 May contains essentially no flux in the NIR from SN 2009ip, with the possible exception of some line emission from He 1.08 $\upmu$m.}
\label{fig:nir}
\end{figure*}
  
The Mg~{\sc i} line at 1.5 $\upmu$m is not observed in either of the NIR spectra, although we note that the S/N per pixel in the continuum is $\sim$3 for the 2013 May 6 spectrum. We take the measured flux of 2.5$\times 10^{-17}$ erg s$^{-1}$ cm$^{-2}$\ in the weakly detected (and nearby) 14$\rightarrow$4 Brackett line as a conservative upper limit to the flux in the Mg line. The measured flux of the possible Mg~{\sc i}] $\lambda$4571 line at this epoch is 1.6$\times 10^{-15}$ erg s$^{-1}$ cm$^{-2}$\, implying that the latter is at least $\sim$60 times stronger. This implies that that the formation of Mg~{\sc i}] $\lambda$4571 and Mg~{\sc i} 1.5 $\upmu$m are different, with the former a cooling line from primordial gas, while the latter is a recombination line with no strong contribution from primordial Mg.

The extreme red part of the XShooter spectrum shows an increase in flux at around 2.25$ \upmu$m and beyond. Fig. \ref{fig:co} shows a comparison of this to the Pfund series in the limit of 5$\rightarrow \infty$, also shown are the bandheads from molecular CO emission. The observed increase in flux is roughly consistent with the position of these, and while the flux calibration of the observational data is least reliable at the edges of spectra, it seems plausible that CO has begun to form by $\sim$280 days after the start of the 2012a event. If this is the case, then the measured 4.5 $\upmu$m fluxes from {\it Spitzer} + IRAC presented in Section \ref{s2} should be regarded as upper limits to the continuum flux, as the filter covers the wavelength of the fundamental emission from CO. We note that CO band heads have been seen in before in cool, evolved stars \citep{Bie02}; while CO has also been seen in Eta Car, where it is believed to have formed in  CNO-processed material \citep{Loi12}. From this, the possible presence of CO emission here cannot be taken as evidence either for or against a core-collapse.

\begin{figure}
\centering
\includegraphics[width=0.7\linewidth,angle=270]{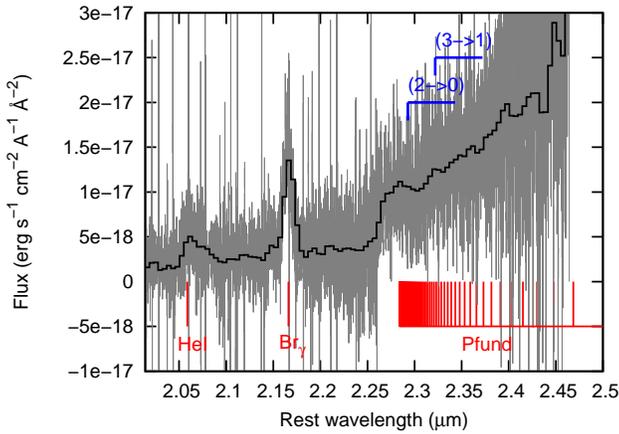}
\caption{{\it K}-band region of the XShooter spectrum from 2013 May 6 (in grey) showing the increasing flux at red wavelengths, along with a smothered, binned version of the same spectrum shown with a black line. The positions of the CO $3\rightarrow2$ and $2\rightarrow0$ band heads are also shown, along with the Pfund series and Br$\gamma$ and He~{\sc i} 2.06$\upmu$m.}
\label{fig:co}
\end{figure}

\section{Discussion and conclusions}
\label{s5}

Despite the fact that SN 2009ip is one of the best monitored extragalactic transients to date, even five years after its initial discovery it continues to present a puzzle. Perhaps the most significant among these questions is whether SN 2009ip has indeed exploded as a core-collapse SN.

\subsection{The energetics of SN 2009ip}

The lightcurve of the 2012a event appears to follow a steady rise to peak, followed by a symmetric decline. This is reminiscent of the diffusion peak seen in Type I SNe. In fact, in the scenario proposed by \cite{Mau13} for SN 2009ip, the 2012a event was the weak explosion of SN 2009ip as a core-collapse SN (while the 2012b event was caused by the ejecta from this event hitting the CSM).

Following \cite{Arn96}, the diffusion timescale for a homologously expanding, spherical body of uniform density can be shown to be approximately
\begin{equation}
t_d = \left(  
3\kappa M_\mathrm{ej}
\over
4 \pi c v_\mathrm{ej}
 \right)^{1/2}
\end{equation}
where $M_\mathrm{ej}$ and $v_\mathrm{ej}$  are the ejecta mass and velocity respectively, $\kappa$ is the opacity of the ejecta, and $c$ is the speed of light. If we take the $\sim$20 day rise of the 2012a event as being characteristic of the diffusion time, and a velocity of 8500 \kms\ (from the minima of the P Cygni absorptions in the Balmer lines; \citealp{Pas13}), and assume an opacity $\kappa$=0.35 cm$^2$~g$^{-1}$ (assuming e$^{-}$ scattering dominates in the H-rich ejecta; \citealp{Ryb79}), the ejecta mass of the 2012a event is only $\sim$0.44 \msun.

We note that there are two caveats to this estimate, namely whether such a simplistic estimate is appropriate here, and whether the absence of photometric coverage in the UV bands during the rise of 2012a event has a significant influence on the bolometric lightcurve. The optical flux after maximum in the 2012a event was a factor $\sim$6 greater than the UV flux \citep{Fra13}, and while the effective temperature prior to the 2012a event maximum was higher, it did not reach the 17000~K seen in the 2012b event when the UV flux contribution {\it did} dominate prior to maximum light (\citeauthor{Fra13}). All in all, we consider it unlikely that the shape of the 2012a event lightcurve would be drastically different with wider wavelength coverage. Whether such a simplistic estimate of the lightcurve is suitable in the case of SN 2009ip is less certain. In particular, the Arnett model assumes a centrally located power source for a transient, while in the case of SN 2009ip, this may be supplemented by additional energy from interaction with a CSM. Non-spherical geometry may also significantly affect the behaviour of SN 2009ip. However, and with these caveats in mind, a toy model may still provide a useful guide as to the energetics and masses involved. In particular, if the 2012a event is the core-collapse of SN 2009ip, then the ejecta mass estimate should be reasonably valid, and at least should provide an upper limit.

If we assume that 0.44 \msun\ was ejected at $\sim$8500 \kms\ at the start of the 2012a event, and that the 2012b event was powered by the kinetic energy of this material, then the $\sim$ 50 days until the start of the 2012b event would imply the inner edge of the CSM lies at 4$\times10^{15}$~cm (270 AU) from the progenitor.

Toy models have been used to estimate the ejecta and CSM characteristics of interacting SNe \citep[e.g.][]{Chu94,Mor13,Che94}. From \cite{Mor13}, we see that in the case of steady mass loss ($\rho \propto R^{-2}$), the late time luminosity decline should follow  $\propto t^{-1.5}$ (eq. 29). In the case of SN 2009ip, we measure a decline with a time dependence $\propto t^{-1.74}$, as shown in Fig. \ref{fig:model}. The disagreement with the expected slope for steady mass loss is unsurprising, given the observed eruptive behaviour of SN 2009ip prior to the 2012a and 2012b events. 

\begin{figure}
\centering
\includegraphics[width=0.7\linewidth,angle=270]{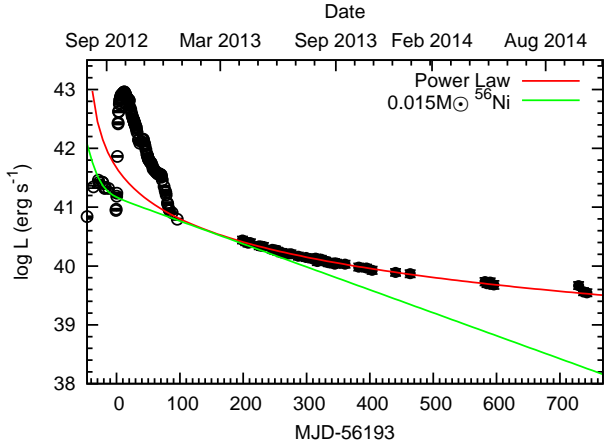}
\caption{Bolometric lightcurve of SN 2009ip, compared to the energy from the decay of 0.015 \msun\ of $^{56}$Co (based on the upper limit to the mass of $^{56}$Co ejected by SN 2009ip from \protect\cite{Fra13}; assuming full trapping of $\gamma$-rays in the ejecta, and to a power law fit to the late time lightcurve with exponent $-$1.74).}
\label{fig:model}
\end{figure}

The emergence of the red shoulder seen in the H$\alpha$ line at late times (shown in more detail in Fig. \ref{fig:halpha_late}) is evidence that there is still structure in the CSM around SN 2009ip at this epoch. A possible formation mechanism for the feature could be interaction between ejecta and CSM on the far side of SN 2009ip, where the energy from X-rays created at the shock interface is absorbed locally; i.e. the emergence of a new kinematical component. In this case, the absence of a corresponding feature in the blue wing of H$\alpha$ indicates asymmetry about SN 2009ip along the line of sight (consistent with the findings of \citealp{Mau14} from spectropolarimetry). We note that a similar line profile was seen for some Type IIn SNe, including SN 1995N \citep{Pas11}. The decrease in prominence of the wings of H$\alpha$ over time is indicative of reduced electron scattering. This is a consistent with the decreasing energy input, giving rise to lower ionization.

\begin{figure}
\centering
\includegraphics[width=1\linewidth,angle=0]{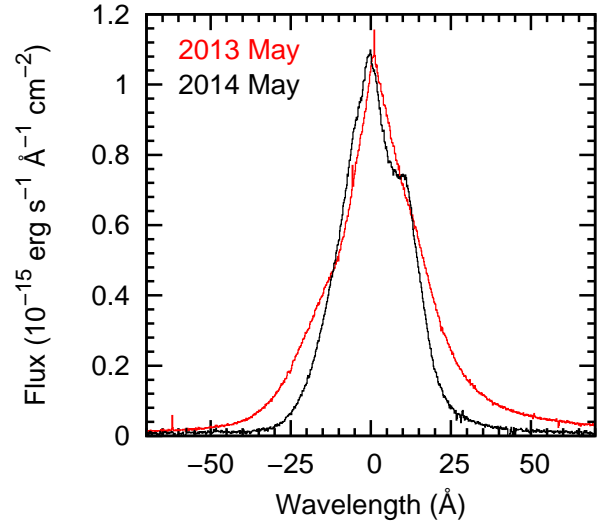}
\caption{The profile of H$\alpha$ as seen in the XShooter spectrum obtained on 2014 May 7 (black), compared to the line profile observed for H$\alpha$ one year earlier on 2013 May 6 (red). The wavelength scale is shown relative to the rest wavelength of the line, the red emission shoulder discussed in the text is clearly visible at +550 \kms.}
\label{fig:halpha_late}
\end{figure}

\subsection{Possible nucleosynthesis in SN 2009ip}

The presence of strong, broad O in late-time, nebular spectra is a classical signature of a core-collapse SN (CCSN). [O~{\sc i}] $\lambda\lambda$6300,6364 is formed from both primordial O in the envelope, and synthesised O from the core of the progenitor, and the doublet is seen to be stronger for more massive progenitors (which contain more massive cores, \citealp[e.g.][]{Jer14}). The appearance of weak [O~{\sc i}] in the spectrum of SN 2009ip approximately a year after the 2012a event could hence be seen as evidence that SN 2009ip did indeed become a genuine SN. However, if this were the case, then we would expect [O~{\sc i}] to become progressively more prominent over time, as the density of the line emitting region in the SN core decreases \citep[e.g.][]{Jer12}. In contrast, we see that the [O~{\sc i}], and indeed the other lines which are plausibly associated with a CCSN such as [Fe~{\sc ii}] $\lambda$7155, remain weak.

We also note that even without nucleosynthesied material from either a SN explosion or stellar evolution prior to collapse, radiative transfer models of nebular Type II SN spectra can show emission from C, Na and O solely arising from primordial material which was present in the stellar envelope. In Fig. \ref{fig:anders}, we plot the emission from the inner H zone at 400 days, from the CCSN model of \cite{Jer12}. This model shows a reasonable match to the observed 2013 October 2 spectrum of SN 2009ip, which was taken at a similar epoch relative to the start of the 2012a event. The [O~{\sc i}] $\lambda\lambda$6300,6364 lines are present in the \cite{Jer12} model, and are actually stronger than those observed in SN 2009ip. In this model, cooling of the H gas occurs through metal lines such as [Ca~{\sc ii}] $\lambda\lambda$7291,7323, the Ca~{\sc ii} NIR triplet, [Fe~{\sc ii}] $\lambda$7155, [O~{\sc i}] $\lambda\lambda$6300,6364, and Mg~{\sc i}] $\lambda$4571, while recombination gives rise to lines from H, He, and some of the metals. The presence of these lines emphasises the need for detailed modelling to distinguish a spectrum arising from H gas with metals at solar abundances (as in a wind or non-terminal outburst) and that of such gas plus additional metals (as in a CCSN).  

From \cite{Gro14}, a 60 \msun\ star (as inferred by \citealp{Fol11} for the progenitor of SN 2009ip from pre-explosion images) will have a final mass of 12 \msun\ at the point of explosion. Such a star will produce a few solar masses of O during hydrostatic burning before it undergoes core-collapse \citep{Woo07b}, which is on the order of the ejected O mass in SN 1987A \citep{Li92}. If we assume that the density in SN 2009ip with respect to SN 1987A will scale according to the inverse cube of their relative ejecta velocities at late times, then the O density in SN 2009ip would be a factor $\sim$10 higher than the latter. As optical depth evolves with $t^{-2}$, one would have to wait $\sqrt{10}$ ($\sim$3) times longer for the transition to optically thin conditions in SN 2009ip. As this transition occurs after a few hundred days in most core-collapse SNe, it may take several years in SN 2009ip. Our latest spectrum of SN 2009ip was taken at $\sim$2 years after its possible core-collapse in August 2012, so the appearance of [O~{\sc i}] lines would not be implausible at this epoch.

The ejected $^{56}$Ni mass from SN 2009ip is another poorly constrained parameter. While energy from CSM interaction contributes to an unknown fraction of the flux in the late-time lightcurve, it is only possible to set an upper limit of the Ni mass. Still, the limit derived (M$_{Ni}<$0.02 \msun) seems lower than would be expected for a $\sim$60 \msun\ progenitor \citep{Nom06} without fallback.

While [O~{\sc i}] is usually associated with CCSNe, there is a precedent for a stellar object displaying these lines, in the case of the B supergiant CPD-52$\degree$9243 \citep{Swi81,Win89}. Interestingly, CPD-52$\degree$9243 is classified as a B[e] star; a class of massive stars which show evidence for a circumstellar disk, reminiscent of the torus around SN 2009ip inferred from spectropolarimetry \citep{Mau14}. We also note that the appearance of [O~{\sc i}] $\lambda\lambda$6300,6364 is not ubiquitous in purported SNe - for example, no strong [O~{\sc i}] was seen in the Type IIn SN 1995G at $\sim$2 years from explosion \citep{Pas02}, while in contrast these lines were clearly seen for SN 1995N \citep{Fra02}. The reason for the absence of the lines in some SNe is unclear; in some cases it may indicate that these are not genuine CCSNe but are rather SN impostors, while in other cases (such as the bright, and slowly declining SN 1995G) it may simply reflect higher densities.

\begin{figure*}
\centering
\includegraphics[width=0.8\linewidth,angle=0]{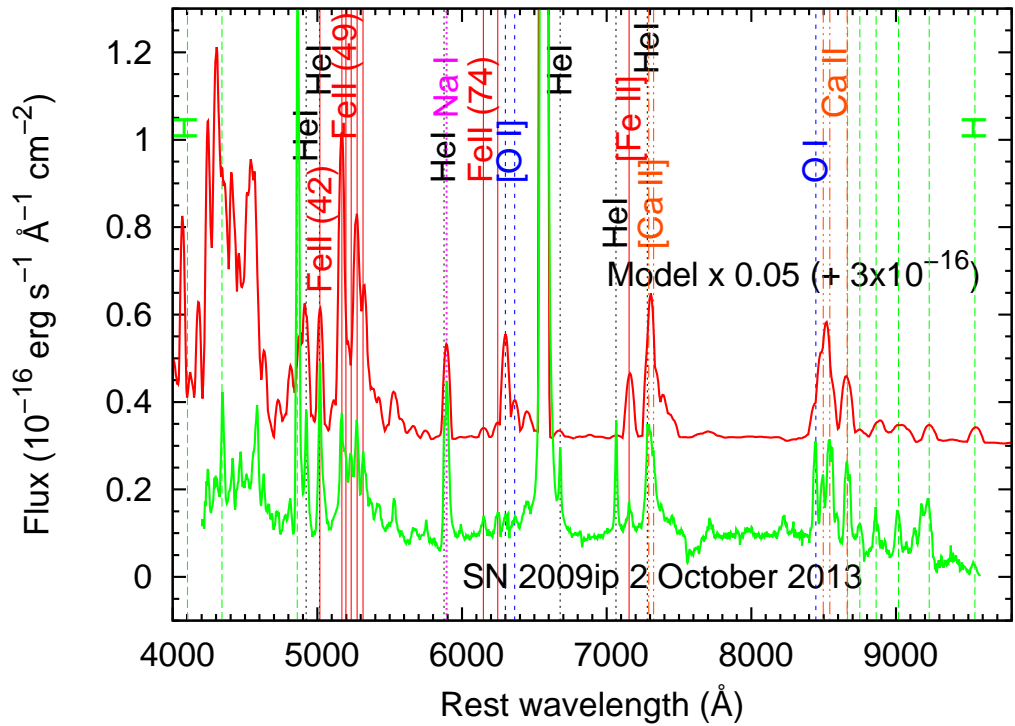}
\caption{The computed emission flux from a model of the inner H zone of a core-collapse SN, showing only line flux arising from material present in the envelope of the pre-explosion progenitor, based on the model of \protect\cite{Jer12} at +400 days. Also shown is the spectrum of SN 2009ip obtained on 2014 October 2, approximately 400 days after the start of the 2012a event.}
\label{fig:anders}
\end{figure*}

\section{Conclusions}

Questions remain as to if, how, and why SN 2009ip exploded. There is now evidence that some very massive stars can explode as Type IIn SNe during a luminous blue variable (LBV) phase \citep[e.g.][]{Tru08,Gal09}. According to stellar evolutionary theory, such massive stars are not expected to explode as H-rich LBVs, but rather are believed to lose their envelopes to become Wolf-Rayet stars, before exploding as Type Ibc SNe. \cite{Gro13,Gro14} have recently used models of massive stars, coupled with a model atmosphere code, to predict the appearance of massive stars shortly prior to collapse. From detailed modelling of a rotating 60 \msun\ star (similar to the progenitor proposed for SN 2009ip by \citealp{Fol11}), they find that it will undergo an LBV phase between the end of core H-burning and the start of core He-burning. The star will ultimately evolve to a WO star, and explode a H-poor SN. While the peripheral location of SN 2009ip within NGC 7259 is suggestive of a low metallicity progenitor, it is unclear whether this could reduce the mass loss rate by enough for a 60 \msun\ star to retain its H envelope to the point of core-collapse. \cite{Heg03} find that while their non-rotating models above $\sim$40 \msun\ result in a H-free SN, at lower metallicities the star will collapse directly to form a black hole, with no accompanying optical display.

Neither the core-collapse nor the non-core-collapse scenarios can be entirely excluded yet for SN 2009ip. The energetics and ejecta velocities seen during the 2012a and 2012b events are closer to those of a core-collapse SN than a SN impostor. On the other hand, broad (albeit much weaker) lines were seen in SN 2009ip as early as 2011 \citep{Pas13}. Furthermore, the low ejected Ni mass is difficult to reconcile with the collapse of a $\sim$60 \msun\ star, unless fallback is invoked. However, even fallback presents problems, as one would presumably have to fine-tune a model to hide nearly all the Ni and core ejecta from SN 2009ip, but still obtain ejecta velocities of $\sim8500$~\kms\ as seen during the 2012a event. It could also be argued that the apparent absence of broad nebular SN lines is evidence against core-collapse, however the density in the ejecta may still be too high for their formation. 

Finally, the physical mechanism powering the outbursts of SN 2009ip prior to 2012 remains unknown. As discussed in \cite{Fra13c} for SN 2011ht, and in \cite{Smi14} for SN 2009ip, the timing of the late time eruptions so soon before a possible core-collapse may suggest a link between the two. The availability of well-sampled pre-discovery light curves for a large sample of SNe can help shed light on the ubiquity of these eruptions \citep{Ofe14}, and help distinguish between some of the proposed explanations \cite[e.g][]{Qua12,Woo07} for the mass loss. Of course, if SN 2009ip can be conclusively demonstrated to have {\it not} undergone core-collapse (for example, if it is seen to undergo a new outburst), then the classification and interpretation of observationally similar transients such as SN 2010mc \citep{Ofe13a} will necessarily need to be revisited.

Continuing observations of SN 2009ip are required to monitor the photometric and spectroscopic evolution of SN 2009ip; and in particular to check if it fades significantly below the magnitude of the progenitor candidate. Alongside this, detailed spectral and hydrodynamic modelling is required to interpret the wealth of observational data obtained so far, to securely identify the lines seen in SN 2009ip, and to form a clearer picture of the true nature of SN 2009ip.

\section{Acknowledgements}

Based on observations collected at the European Organisation for Astronomical Research in the Southern Hemisphere, Chile as part of ESO programs 291.D-5010 and 092.D-0586, and as part of PESSTO, (the Public ESO Spectroscopic Survey for Transient Objects Survey), ESO program 188.D-3003/191.D-0935. This work is based in part on observations made with the Spitzer Space Telescope, which is operated by the Jet Propulsion Laboratory, California Institute of Technology under a contract with NASA (Proposal ID 90255). Based on observations (GS-2013A-DD-3 and GS-2012B-Q-86) obtained at the Gemini Observatory, which is operated by the Association of Universities for Research in Astronomy, Inc., under a cooperative agreement with the NSF on behalf of the Gemini partnership: the National Science Foundation (United States), the National Research Council (Canada), CONICYT (Chile), the Australian Research Council (Australia), Minist{\'e}rio da Ci\r{e}ncia, Tecnologia e Inova\c{c}{\~a}o (Brazil) and Ministerio de Ciencia, Tecnolog{\'i}a e Innovaci{\'o}n Productiva (Argentina). The Liverpool Telescope is operated on the island of La Palma by Liverpool John Moores University in the Spanish Observatorio del Roque de los Muchachos of the Instituto de Astrofisica de Canarias with financial support from the UK Science and Technology Facilities Council. This work makes use of observations from the LCOGT network. Some of the data presented herein were obtained at the W.M. Keck Observatory, which is operated as a scientific partnership among the California Institute of Technology, the University of California and the National Aeronautics and Space Administration. The Observatory was made possible by the generous financial support of the W.M. Keck Foundation.  The authors wish to recognize and acknowledge the very significant cultural role and reverence that the summit of Mauna Kea has always had within the indigenous Hawaiian community.  We are most fortunate to have the opportunity to conduct observations from this mountain.

We thank the directors of the Gemini Observatory and ESO for their crucial allocations of discretionary time to this project.  We thank Emma Reilly, Jon Mauerhan, Nathan Smith, Ori Fox and Raffaella Margutti for helpful discussions at the CAASTRO ``Supernova 2014'' workshop.

This work was partly supported by the European Union FP7 programme through ERC grant number 320360.
R.K. acknowledges funding from STFC (ST/L000709/1).
Part of this  research was conducted by the Australian Research Council Centre of Excellence for All-sky Astrophysics (CAASTRO), through project number CE110001020.
S.J.S acknowledges funding from the European Research Council under the European Union's Seventh Framework Programme (FP7/2007-2013)/ERC Grant agreement n$^{\rm o}$ [291222] and STFC grants ST/I001123/1 and ST/L000709/1. 
A.M.G. acknowledges financial support by the Spanish Ministerio de Econom\'ia y 
competitividad (MINECO) grant ESP2013-41268-R.A.G.-Y. is supported by an EU/FP7-ERC grant no [307260], ``The Quantum Universe'' I-Core program by the Israeli Committee for planning and budgeting and the ISF, GIF, Minerva, ISF and Weizmann-UK grants, and the Kimmel award.
F.E.B acknowledges support from CONICYT-Chile (Basal-CATA PFB-06/2007, FONDECYT 1141218, PCCI 130074) and Project IC120009 "Millennium Institute of Astrophysics (MAS)" funded by the Iniciativa Cient\'{\i}fica Milenio del Ministerio de Econom\'{\i}a, Fomento y Turismo.
S.H. is funded from a Minerva ARCHES award of the German Ministry of Education and Research (BMBF).
A.P. and S.B. are partially supported by the PRIN-INAF 2011 with the project ``Transient Universe: from ESO Large to PESSTO''.
N.E.R. acknowledges the support from the European Union Seventh Framework Programme (FP7/2007-2013) under grant agreement n. 267251 ``Astronomy Fellowships in Italy'' (AstroFIt).

\end{document}